\documentclass[fleqn,usenatbib]{mnras}

\usepackage{newtxtext,newtxmath}

\usepackage[T1]{fontenc}

\DeclareRobustCommand{\VAN}[3]{#2}
\let\VANthebibliography\thebibliography
\def\thebibliography{\DeclareRobustCommand{\VAN}[3]{##3}\VANthebibliography}

\usepackage{graphicx}	
\usepackage{amsmath}	
\usepackage{comment}
\usepackage[utf8]{inputenc}
\usepackage{makecell}
\usepackage{hyperref}
\usepackage{verbatim}
\usepackage{pbox}
\usepackage{physics}
\usepackage{amsmath}

\newcommand{\vect}[1]{\boldsymbol{#1}}

\title[NIR interferometry of A stars]{Binarity and beyond in A stars \\ I. Survey description and first results of VLTI/GRAVITY observations of VAST targets with high \textit{Gaia-Hipparcos} accelerations\thanks{Based on observations collected at the European Southern Observatory, Chile, Program IDs 105.20RL.001, 105.20RL.002, 109.23BV.001, 109.23BV.002}}

\author[Waisberg, Klein \& Katz]{
Idel Waisberg,$^{1}$\thanks{E-mail: idel.waisberg@weizmann.ac.il}
Ygal Klein,$^{1}$
and Boaz Katz$^{1}$
\\
$^{1}$Department of Particle Physics and Astrophysics, Weizmann Institute of Science, Rehovot 76100, Israel
}

\date{Accepted XXX. Received YYY; in original form ZZZ}

\pubyear{2022}

\begin{document}
\label{firstpage}
\pagerange{\pageref{firstpage}--\pageref{lastpage}}
\maketitle

\begin{abstract}
A-stars are the progenitors of about half of the white dwarfs (WDs) that currently exist. The connection between the multiplicity of A-stars and that of WDs is not known and the observational mapping of both multiplicities are far from complete. Possible companions at separations of tens of AU are particularly poorly explored. We are conducting a near-infrared interferometric survey with VLTI/GRAVITY of twenty out of 108 southern A stars within the VAST sample which show large \textit{Gaia-Hipparcos} proper motion changes suggestive of a $M \sim 1 M_{\odot}$ companion at separations of $1-20$ AU. 
In this paper, we detail our sample selection and report on the interferometric detection of $8_{-0}^{+2}$ new stars (including four high multiplicity (3+) systems) in a partial sample of 13 targets. Moreover, we also conduct a common proper motion search for the 108 A stars using \textit{Gaia} eDR3 and which resulted in 10 new detections and confirmation of several previous Adaptive Optics companions as physical. We discuss our preliminary results in the context of the separation distribution of A stars and implications for the multiplicity of WDs. In particular, we find that (i) the apparent suppression of companions to A stars below about 30-50 AU is very likely due to an observational bias, (ii) the fact that 4 of the 6 closest WDs have a companion within a few tens of AU is a statistical fluke but 10-20 such binaries are likely still missing within 20 pc, (iii) a large fraction of such systems likely had high multiplicity (3+) progenitors with very close ($< 1$ AU) companions to the primary A star, and must therefore have undergone non-trivial evolution. 
\end{abstract}




\section{Introduction}
The assessment of the multiplicity of nearby stars presents a long standing observational challenge motivated by the implications of multiplicity to our understanding of stellar formation and evolution  \citep[e.g.][]{Abt76,DuquennoyMayor91,Raghavan10,DeRosa14}. White Dwarfs (WDs) are the end states of Main Sequence (MS) in the mass range $1 M_{\odot} - 8 M_{\odot}$ and the study of the multiplicity of both these classes presents a unique opportunity to probe the evolution of such systems \citep[e.g.][]{Toonen17} and in particular their suspected connection to type Ia Supernovae \citep[e.g.][]{Hillebrandt00,Maoz14} . A-type stars, with masses in the range of $1.5-2.5 M_{\odot}$, evolve to become white dwarfs (WDs) with masses in the vicinity of $0.6 M_{\odot}$ \citep[e.g.][]{Catalan08,Cummings18}, where the WD mass distribution is peaked \citep[e.g.][]{Kepler19}. They consist of an important subgroup of WD progenitors which on the one hand represents about half of the progenitors while on the other hand is relatively abundant and homogeneous compared to the entire range.  

An important recent advancement on A star multiplicity was made with the VAST survey \citep{DeRosa14}, which collected adaptive optics (AO) observations on a sample of 363 photometrically selected A stars within 75 pc and provided high completeness for massive companions ($M> 1 M_{\odot}$) with projected separations $\rho \gtrsim 0.3''$ ($a \gtrsim 12-20 \text{ AU}$ for distances $40 \text{ pc} \lesssim d \lesssim 75 \text{ pc}$). \cite{DeRosa14} also provided a common proper motion (CPM) search for wide companions, which has in the meantime been superseded by \textit{Gaia} \citep{Gaia16,Gaia18,Gaia21}. Companions with separations $a \lesssim 1.5 \text{ AU}$ (corresponding to orbital periods $P \lesssim 1 \text{yr}$ for a total mass $M=3 M_{\odot}$) can in principle be revealed by photometric \citep[e.g.][]{Murphy18} or radial velocity (RV) variations, although such observations are demanding and deriving the RV curves is challenging if many components are present in the spectrum \citep{Gullikson16}. Currently the separation range $20-40 \text{ mas} \lesssim \rho \lesssim 300 \text{ mas}$ ($2 \text{ AU} \lesssim a \lesssim 12-20 \text{ AU}$ for $40 \text{ pc} \lesssim d \lesssim 75 \text{ pc}$) remains largely unexplored.

A somewhat surprising result from the VAST survey is a significant suppression of companions to A stars below 30-50 AU, with a reduction by a factor of around 3 at 30 AU (the lower limit at which the survey is complete to low mass companions). The distribution is known to rise up again to comparable levels at short separations $\sim 1-3 \text{ AU}$ based on dedicated spectroscopic surveys \citep[e.g.][]{Abt65}. Taken together, there seems to be a significant gap in the distribution of companion separations around tens of AU. Is this gap real or is this an artifact of the limited sensitivity for companions at this range?

There are reasons to expect that companions at separations of a few tens of AU are actually common. In particular, 4 out of the 6 WDs within 6 pc (Sirius B, Procyon B, 40 Eridani B and Stein 2051 B) have MS companions with projected separations $10$ AU $< a <60$ AU, implying progenitor systems consisting of MS-MS intermediate mass binaries with $1$ AU $\lesssim a \lesssim 30$ AU separations. There are, however, very few such systems known in this separation range at larger distances from the Sun \citep[][see Fig. \ref{fig:WD_20pc}]{DeRosa14,Holberg13,Hollands18} indicating that many such systems are missed \citep[e.g.][]{Katz14} and inhibiting a conclusive estimate of their abundance. 

The companions of the closest two WDs (Sirius A and Procyon A) are sufficiently massive ($M> 1 M_{\odot}$) to evolve within a few Gyrs implying that these systems will become WD-WD binaries presenting another puzzle that may be related to our poor sensitivity to companions at tens of AU. It is expected that about 15-20 percent of WDs should have WD companions \citep[][]{Klein17,Toonen17} and yet only two are observed within the 20 pc sample which consists of 140 WDs and is claimed to be almost complete \citep[][]{Holberg08,Hollands18}. 

With the goal of better characterizing the multiplicity of A stars, in particular for $M \sim 1 M_{\odot}$ companions at separations of tens of AU, we are in the process of conducting a small NIR survey of 20 A stars in the southern hemisphere which are part of the VAST survey and which show a large change in \textit{Gaia-Hipparcos} proper motion \citep{Brandt18,Brandt21}. The goal of the interferometry is not only to explore the $20-30 \text{ mas} \lesssim \rho \lesssim 300 \text{ mas}$ seaparation range mentioned above, but also to find close companions in high multiplicity systems which may have escaped detection due to the complexity of their spectrum. Such systems are particularly interesting because multiplicity beyond binarity may play an important role in the evolution of the system before, during or after stellar evolution \citep[e.g.][]{Harrington68,Kiseleva98,Tokovinin06,Fabrycky07,Perets09,Toonen16,Hamers21, Gao22} and may affect the prospects of the system to result in a type Ia supernovae \citep[e.g.][]{Thompson11,Katz12,Kushnir13}.  

In this paper, we present the results of the observations of a partial sample of the 20 A stars we are targeting. This paper is organized as follows. In Section 2, we describe our strategy in selecting promising targets for NIR interferometry follow-up based on their \textit{Gaia} and \textit{Hipparcos} astrometry. In Section 3, we describe the VLTI/GRAVITY observations and data reduction. In Section 4, we describe the details of the interferometric modeling to find or constrain companions, as well as the details of our search for common proper motion companions with \textit{Gaia}. In Section 5, we provide background and show the results for each individual object observed with interferometry, while on Section 6 we summarize our results for the CPM search. Finally, in Section 7 we discuss our results in the context of multiplicity of A stars and their WD descendants. 

\section{Sample strategy}

We select our parent sample of 108 A stars by selecting those stars in the VAST sample of 435 A stars \citep{DeRosa14} which have (i) $\text{DEC}<0$ (150 stars), because VLTI/GRAVITY is located in the southern hemisphere, (ii) previous AO observations (124 stars), and (iii) distance $40 \text{ pc} \lesssim d \lesssim 75 \text{ pc}$ (108 stars), to ensure a better uniformity in terms of physical separations probed for the whole sample. We note that the VAST sample is not complete, particularly in the South (there are 636 photometrically selected A stars within 75 pc but 201 of them were not part of the survey, most of them in the South); however, as far as we are aware, there is no bias in the selection of the stars that were part of the survey. Out of our 108 A stars, 7 are classified as metallic-line (Am) and only 3 as chemically peculiar (Ap) in the SIMBAD database.

Next, we use \textit{Gaia} and \textit{Hipparcos} astrometry in order to select promising targets. Specifically, we use the difference between the \textit{Gaia} DR2 \citep{Gaia18} proper motion and the proper motion inferred by the positional difference between \textit{Gaia} DR2 and \textit{Hipparcos} \citep{Brandt18}. These two proper motions are effectively separated by 12 years, and the reason to use the latter instead of the actual \textit{Hipparcos} proper motion is because it is much more precise. If we consider a companion perturber with mass $M_{\mathrm{per}}$ at a projected separation $a_{\mathrm{proj}}$, the proper motion change it would cause on the A star over $\Delta t = 12 \text{yrs}$ is 

\begin{align}
\Delta v_{\mathrm{proj}} = \frac{G M_{\mathrm{per}}}{a_{\mathrm{proj}}^2} \Delta t \sin^3\alpha \\
=5.6 \text{ km}\text{ s}^{-1}  \frac{M_{\mathrm{per}}}{1 M_{\odot}} \left ( \frac{a_{\mathrm{proj}}}{20 \text{ AU}} \right )^{-2} \frac{\Delta t}{12 \text{ yrs}} \sin^3\alpha
\end{align}

\noindent where $\alpha$ is the angle between the line connecting the two components and the direction to the observer and $\sin^3 \alpha > 0.08$ for $90\%$ of orientations, and this is valid for orbital periods somewhat longer than $\Delta t$ (e.g. $P \gtrsim 30 \text{ yrs}$ for $\Delta t = 12 \text{ yrs}$). Based on this, we select the targets among the 108 A stars that have $\Delta v_{\mathrm{proj}} > 0.5 \text{ km} \text{ s}^{-1}$ for interferometric follow-up, which gives us essentially full completeness for $M_{\mathrm{per}} > 1 M_{\odot}$ and $a_{\mathrm{proj}} < 20 \text{ AU}$. While this will certainly include any $M_{\mathrm{per}} > 1 M_{\odot}$ within such separation range, it will also include less massive companions on closer orbits or with favorable orientations. We note that we have assumed that the light centroid is the A star, which is a very good approximation as long as the mass ratio is not too close to unity given the steep mass-luminosity relation for MS stars (in the extreme case of a twin binary there is no proper motion change). For periods smaller than about 30 yrs, a substantial fraction of the orbit is covered within the baseline and the velocity change is limited by the orbital velocity. In this case, the proper motion difference is roughly proportional to the orbital velocity and therefore to $a^{-1/2}$.  

For very short periods $P \ll 1 \text{ yr}$ the orbital motion averages out in the \textit{Hipparcos} and \text{Gaia} proper motion measurements, although such orbits might be reflected in other parameters such as the astrometric noise and the Renormalized Unit Weight Error provided by \textit{Gaia} eDR3 \citep[RUWE;][]{Belokurov20}. Since some care is needed in the interpretation of this value for any given individual target \citep{Lindegren21}, we do not use these parameters in our sample selection (their use to identify short period spectroscopic binaries is beyond the scope of this paper). On the other hand, for periods $P \gtrsim 0.5 \text{yr}$, the orbital motion might not average out in the \textit{Gaia} proper motion measurement depending on the orbit and the sampling. 
    
\begin{figure*}
 \includegraphics[width=2\columnwidth]{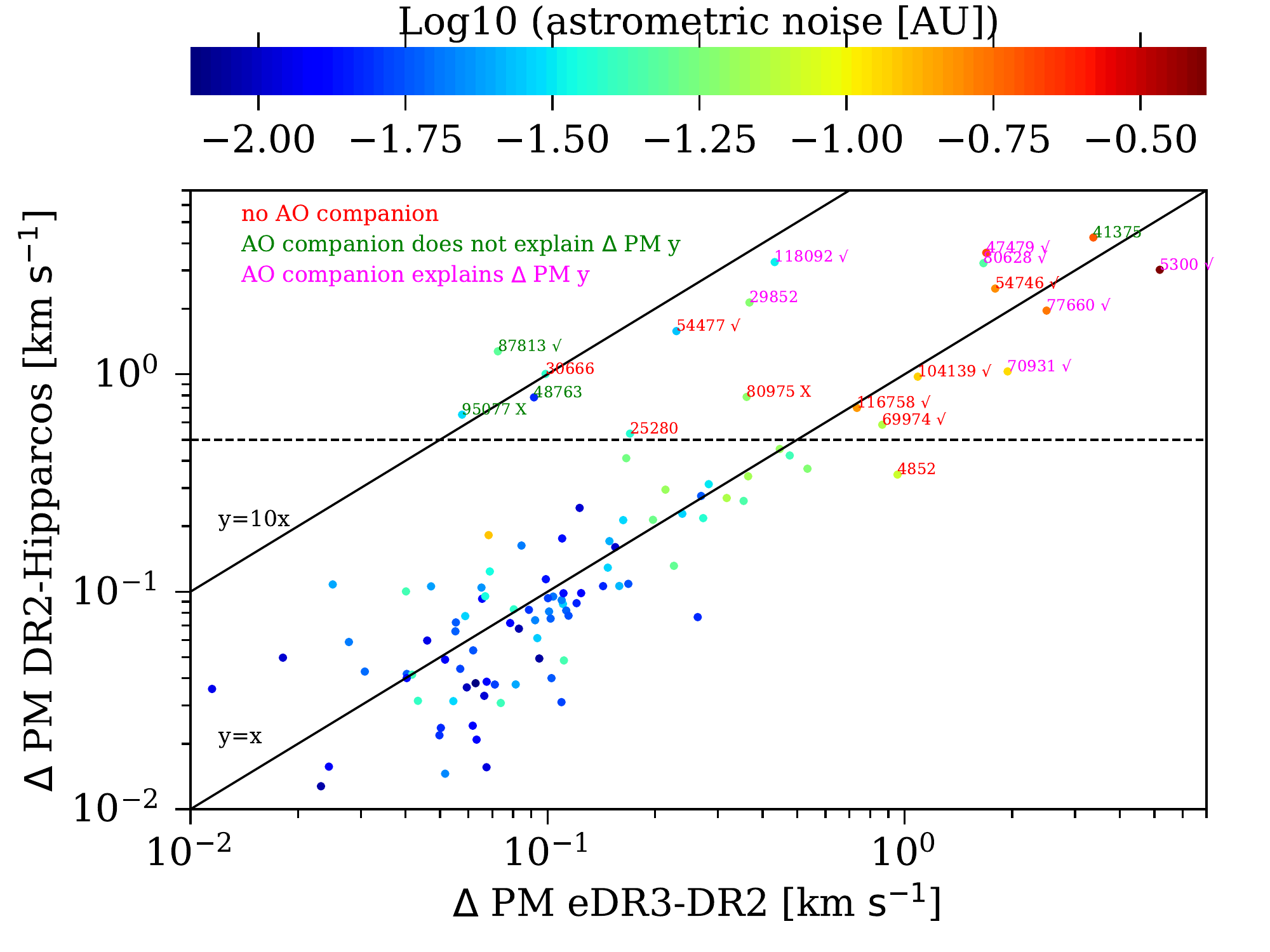}
 \caption{\textit{Hipparcos}-\textit{Gaia} DR2-\textit{Gaia} eDR3 plane for our parent sample of 108 A stars. The change in proper motion between \textit{Gaia} DR2 and that inferred from the \textit{Gaia}-\textit{Hipparcos} change in position  \citep[][mean epochs of these proper motion baselines differ by 12 years]{Brandt18} is shown vs the change in proper motion between \textit{Gaia} eDR3 and DR2 for the entire sample (mean epochs differ by 0.5 yr). The 20 A stars which form our interferometric sample are labeled by their HIP index. Targets with additional close companions found previously or here are marked by a check-mark. The two targets that were observed here for which no companion was found are marked by $X$. Colors of the symbols indicate \textit{Gaia} astrometric noise (see color-bar at the top of the figure). Colors of the HIP index labels indicate whether or not there is a companion found by AO in VAST and whether or not it can explain the \textit{Gaia}-\textit{Hipparcos} proper motion change on the y-axis of the figure.}
 \label{fig:Hip-Gaia}
\end{figure*}

We can have a rough idea about possible scenarios for a given target by plotting it in what we will call the \textit{Hipparcos}-\textit{Gaia} DR2-\textit{Gaia} eDR3 plane, shown in Figure \ref{fig:Hip-Gaia} for the 108 A stars in our parent sample. In this plot, the y-axis shows the $\Delta v_{\mathrm{proj}}$ described above and the x-axis shows the proper motion difference between \textit{Gaia} eDR3 and DR2, which have mean epochs that differ by 0.5 yr (2015.5 vs 2016.0). The 20 A stars which form our interferometric sample are labeled. \footnote{We point out three particular cases (i) HIP 4852 is part of our sample despite having $y < 0.5 \text{ km} \text{ s}^{-1}$ in Figure \ref{fig:Hip-Gaia} because its y-value is above the threshold when using \textit{Gaia} eDR3 instead of DR2; (ii) HIP 69974 ($\lambda$ Virginis) is labeled in the figure but is not in our sample because it has already been observed with NIR interferometry \citep{Zhao07}; (iii) HIP 47204 is part of our sample but is not shown in Fig. \ref{fig:Hip-Gaia} because \textit{Gaia} does not recover an astrometric solution for it. We included it in our sample because it is a known AO binary with a close massive companion \citep{DeRosa14}.} Their labels are also colored by whether they have an AO companion in \cite{DeRosa14} and if so, whether it can explain the change in the y-axis proper motion (i.e. whether it is massive/close enough). The points are colored according to their astrometric excess noise (in AU) from \textit{Gaia} eDR3. 

In Figure \ref{fig:Hip-Gaia} we expect stars that are near the upper branch (labeled y=10x to indicate that the \text{Gaia} DR2-\textit{Hipparcos} proper motion change is ten times higher than the \textit{Gaia} eDR3-DR2 change) to have a companion on a long period orbit ($P \gtrsim 30 \text{ yrs}$). Two of the stars in this branch already have such an AO companion (HIP 118092 and HIP 29852). Furthermore, we already reported the discovery of such a companion for one of the other targets, HIP 87813, in paper II of this series \citep{Waisberg22}.

On the other hand, for stars that fall near the lower branch (y=x) there are a couple of possibilities. They can be binaries with period $P \gtrsim 1 \text{ yr}$ so that the orbital motion does not average out in the \textit{Gaia} proper motion measurement. This is probably the case for the known spectroscopic binaries HIP 69974 \citep[P=207 days;][]{Zhao07} and HIP 77660 \citep[P=1073 days][]{Nikolov08} -- although the latter also has a massive AO companion on a much longer period orbit which contributes to the DR2-\textit{Hipparcos} difference --, and for HIP 47479, which has an AO companion on a relatively short P=10.74 yrs orbit \citep{DeRosa12}. On the other hand, for systems such as HIP 5300 and HIP 80628, which have AO companions on long period orbits \citep[P=28.36 yrs and P=82.8 yrs, respectively;][]{DeRosa12}, their unexpected location in the lower branch is suggestive of additional companions; our interferometric observations described below indeed show that is the case. We note that the stars with high acceleration in the lower branch consistently have the highest astrometric noise in the entire sample. 

In this paper, we present the interferometric results for a sub-sample of targets that we have already observed. In Table \ref{table:targets} we summarize their basic properties that will be further used in our analysis: 2MASS K band magnitude, Tycho V band magnitude and distance. 
    
\begin{table}
\centering
\caption{\label{table:targets} Properties of the interferometric targets discussed in this paper.}
\begin{tabular}{cccc}
\hline \hline
target & V$_T$ & K & \shortstack{distance$^a$\\(pc)} \\[0.3cm]

HIP 5300 & 5.228 & 4.777 & $70.2_{66.2}^{73.9}$ \\[0.3cm]

HIP 47479 & 5.332 & 4.798 & $68.0_{66.7}^{69.6}$ \\[0.3cm]

HIP 54477 & 5.430 & 5.235 & $54.5_{54.2}^{54.7}$ \\[0.3cm]

HIP 54746 & 5.388 & 4.971 & $54.6_{53.8}^{55.7}$ \\[0.3cm]

HIP 70931 & 5.382 & 5.197 & $60.7_{59.7}^{61.9}$ \\[0.3cm]

HIP 77660 & 5.106 & 4.699 & $57.5_{55.9}^{59.6}$ \\[0.3cm]

HIP 80628 & 4.653 & 4.165 & $39.9_{39.6} ^{40.3}$ \\[0.3cm]

HIP 80975 & 4.449 & 4.137 & $51.3_{50.9}^{51.8}$ \\[0.3cm]

HIP 87813 & 5.934 & 5.698 & $78.2_{77.6}^{78.8}$ \\[0.3cm]

HIP 95077 & 5.614 & 4.922 & $56.6_{56.3}^{57.0}$ \\[0.3cm]

HIP 104139 & 4.070 & 4.100 & $45.7_{44.8}^{46.4}$ \\[0.3cm]

HIP 116758 & 5.008 & 4.342 & $39.9_{39.0}^{40.9}$ \\[0.3cm]

HIP 118092 & 5.957 & 5.703 & $63.8_{63.5}^{64.0}$ \\[0.3cm]
\hline
\multicolumn{4}{l}{$^a$ geometric distance and 1$\sigma$ errors from \cite{Bailer-Jones21}}
\end{tabular}
\end{table}

\section{OBSERVATIONS AND DATA REDUCTION}

\subsection{VLTI/GRAVITY}

The stars were observed with the beam combiner instrument GRAVITY \citep{GRAVITY17} on the VLTI using the 1.8-m Auxiliary Telescopes (ATs). A summary of the observations, including the seeing and coherence time during the observations, is provided in Table \ref{table:obs}. GRAVITY operates in the NIR K band and most of the observations were made in single field mode: half of the light from the object was directed to the fringe tracker (FT), which operates in low spectral resolution mode ($R\approx22$) in order to track and stabilize the fringes, while the other half was directed to the science (SC) channel, where coherent integrations of DIT=30s were made in the high spectral resolution ($R\approx4000$) mode. For one of the targets, HIP 80628, an additional observation was performed in dual field mode: the primary was used as the FT source and the secondary (separated from the primary by about 1") was observed in the SC channel. The observations were done in combined polarization mode and made use of the adaptive optics (AO) system NAOMI \citep{Woillez19} using the optical light from the FT object itself. The total integration time on source was 16 min (separated in four files) for all the sources, except for HIP 47479, for which only a 5 min individual file was obtained. The observations provided spectrally resolved squared visibilities and closure phases across the K band (1.97-2.40$\mu$m). The AT configuration for each observation, as well as the canonical spatial resolution $\theta = \frac{2.2 \mu\text{m}}{B_{\mathrm{proj,max}}}$ corresponding to the largest projected baseline $B_{\mathrm{proj,max}}$, are noted in Table \ref{table:obs} (with the caveat that sources can be partially resolved to small fractions of this value). Each target observation was preceded or succeeded by the observation of a nearby interferometric calibrator, listed in Table \ref{table:obs} and which is used for absolute calibration of the squared visibilities and closure phases. The calibrator properties are listed in Table \ref{table:calibrators}. 

\begin{table*}
\centering
\caption{\label{table:obs} Summary of GRAVITY observations.}
\begin{tabular}{ccccccc}
\hline \hline
target & \shortstack{date\\MJD} & \shortstack{seeing\\@ 500 nm (")} & \shortstack{coherence time\\@ 500 nm (ms)} & \shortstack{AT configuration} & \shortstack{$B_{\mathrm{proj,max}}$ (m) \\$\theta_{\mathrm{max}}$ (mas)} & calibrator \\ [0.3cm]

HIP 5300 & \shortstack{2021-08-07\\59433.34} & 0.6-0.7 & 4 & \shortstack{K0-G2-D0-J3\\medium} & \shortstack{102.3\\4.4} & HD 6571 \\ [0.3cm]

HIP 47479 & \shortstack{2019-03-07\\58549.02} & 0.5-0.6 & 6 & \shortstack{A0-G1-J2-J3\\large} & \shortstack{130.6\\3.5} & HD 106983 \\ [0.3cm]

HIP 54477 & \shortstack{2022-04-09\\59678.13} & 0.9-1.2 & 2-3 & \shortstack{A0-G1-J2-J3\\large} & \shortstack{132.0\\3.4} & HD 98300 \\ [0.3cm]

HIP 54746 & \shortstack{2022-04-02\\59671.11} & 0.7-0.8 & 3 & \shortstack{A0-G1-J2-J3\\large} & \shortstack{128.8\\3.5} & HD 98037 \\ [0.3cm]

HIP 70931 & \shortstack{2022-04-12\\59681.26} & 1.4-1.7 & 2 & \shortstack{A0-G1-J2-K0\\astrometric} & \shortstack{129.0\\3.5} & HD 141072 \\ [0.3cm]

HIP 77660 & \shortstack{2022-04-12\\59681.30} & 1.4-1.8 & 2 & \shortstack{A0-G1-J2-K0\\astrometric} & \shortstack{128.4\\3.5} & HD 141072 \\ [0.3cm]

HIP 80628 & \shortstack{2021-08-23\\59449.01} & 0.6-0.7 & 7 & \shortstack{A0-G1-J2-K0\\astrometric} & \shortstack{123.8\\3.7} & HD 149383 \\ [0.3cm]

HIP 80628 & \shortstack{2022-05-09\\59708.32} & 1.7-2.1 & 1.8-2.1 & \shortstack{A0-G1-J2-J3\\large} & \shortstack{131.0\\3.5} & HD 146949 \\ [0.3cm]

HIP 80975 & \shortstack{2021-08-31\\59457.11} & 0.9-1.2 & 3 & \shortstack{A0-B2-D0-C1\\small} & \shortstack{29.3\\15.5} & HD 149934 \\ [0.3cm]

HIP 87813 & \shortstack{2021-09-02\\59459.10} & 0.7-0.8 & 3-5 & \shortstack{A0-B2-D0-C1\\small} & \shortstack{33.0\\13.7} & HD 163449 \\ [0.3cm]

HIP 95077 & \shortstack{2021-09-02\\59459.16} & 0.7-0.9 & 2-4 & \shortstack{A0-B2-D0-C1\\small} & \shortstack{32.4\\14.0} & HD 182481 \\ [0.3cm]

HIP 104139 & \shortstack{2021-08-10\\59436.22} & 0.6-0.7 & 3-4 & \shortstack{K0-G2-D0-J3\\medium} & \shortstack{104.2\\4.3} & HD 201335 \\ [0.3cm]

HIP 116758 & \shortstack{2021-09-01\\59458.24} & 0.5-0.7 & 3-4 & \shortstack{A0-B2-D0-C1\\small} & \shortstack{33.6\\13.5} & HD 222125 \\ [0.3cm]

HIP 118092 & \shortstack{2021-08-07\\59433.31} & 0.6-0.9 & 3 & \shortstack{K0-G2-D0-J3\\medium} & \shortstack{93.9\\4.8} & HD 224472 \\ [0.3cm]
\hline
\end{tabular}
\end{table*}

\begin{table}
\centering
\caption{\label{table:calibrators} Interferometric calibrators.}
\begin{tabular}{ccccc}
\hline \hline
calibrator & spectral type & V & K & $\theta_{K}^a$ (mas) \\[0.3cm]

HD 6571 & K3III & 7.71 & 4.34 & 0.727 \\[0.3cm]

HD 98037 & K0III & 7.27 & 4.83 & 0.493 \\[0.3cm]

HD 98300 & K1III & 7.71 & 5.19 & 0.439 \\[0.3cm]

HD 106983 & B2/3 V & 4.05 & 4.54 & 0.259 \\[0.3cm]


HD 141072 & K3III & 8.20 & 4.83 & 0.556 \\[0.3cm]

HD 149934 & K0III & 6.98 & 3.36 & 1.099 \\[0.3cm]

HD 163449 & K0III & 4.70 & 7.61 & 0.560 \\[0.3cm]

HD 201335 & K4III & 6.69 & 3.19 & 1.173 \\[0.3cm]

HD 222125 & K0III & 6.38 & 4.24 & 0.643 \\[0.3cm]

HD 224472 & G8IV & 6.86 & 4.80 & 0.505 \\[0.3cm]

\hline
\multicolumn{5}{p{0.4\textwidth}}{$^a$Angular diameter from the \newline JMMC Stellar Diameter Catalog v2 \citep{Bourges17}.}
\end{tabular}
\end{table}

GRAVITY also records a $80 \text{ mas} \text{ pixel}^{-1}$ H band image with a field-of-view of around 4" around the target every second for each telescope with its acquisition camera. This image is very useful scientifically as it has a spatial resolution close to the diffraction limit of the ATs in the H band (250 mas). Unfortunately, some of the targets for which the apparent magnitude $H>5$ were saturated because the attenuation filter by default is only used for $H<5$ (this was corrected in later observations), so that in these cases the effective spatial resolution is significantly reduced due to saturation.

The GRAVITY data were reduced with the GRAVITY instrument pipeline v1.4.0 \citep{Lapeyrere14} downloaded from the ESO website with default options.

\section{Methods} 

\subsection{Multiple star interferometric model}

Our basic model is that of a collection of $N$ point sources, with a complex visibility \citep[e.g.][]{Waisberg19}: 
\begin{equation}
\label{eqn:multiple}
V = \frac{1 + \sum \limits_{i=2}^N \frac{f_{c,i}}{f_{c,1}} f_i e^{-2 \pi j \vect{u}\cdot \vect{\sigma_i}}}{1 + \sum \limits_{i=2}^N \frac{f_{c,i}}{f_{c,1}} f_i}
\end{equation}

\noindent where $0<f_i<1$ is the intrinsic flux ratio between component $i$ and the primary ($i=1$), $\vect{\sigma_i}$ is the separation vector between component $i$ and the primary and $\vect{u}$ is the $uv$ coordinate. $f_{c,i}$ is a fiber coupling factor that takes into account the attenuation of the source away from the center of the GRAVITY fiber. We model it as

\begin{equation}
f_{c,i} = e^{-|d_i|^2/(2 \sigma_{\mathrm{fiber}}^2)}
\end{equation}

\noindent where $d_i$ is the distance of component $i$ from the center of the fiber and $\sigma_{\mathrm{fiber}} = \mathrm{FWHM}_{\mathrm{fiber}}/2.355$ is the standard deviation of the fiber mode, which in the case of GRAVITY is matched to the Point Spread Function (PSF) of the telescope ($\mathrm{FWHM}_{\mathrm{fiber}} = 250 \text{ mas}$ in the case of the 1.8m ATs). For a compact collection of sources ($\lesssim$ a few tens of mas), the fiber coupling factor is a very small correction and is usually ignored; however, for large separations this factor should be taken into account in order to derive accurate flux ratios. We take the center of the fiber to be the light centroid of the system in case there is no clearly resolved companion in the acquisition camera image.

We also note that in Eq. \ref{eqn:multiple}:

\begin{enumerate}
\item We treat the stars as point sources because their size is much smaller than the interferometer resolution. For example, an A star with radius $R=2R_{\odot}$ has an angular diameter $\theta \approx0.4 \text{ mas}$ at a distance of $d=40 \text{ pc}$, which is ten times below the maximum resolution achieved; 

\item We neglect the effect of bandwidth smearing, which reduces the visibility for sources far from the center of the interferometric field of view $\mathrm{FOV}_{\mathrm{interf}} \sim 0.6 \frac{\lambda R}{B}$, where $R$ is the spectral resolution and $B$ the projected baseline. For our GRAVITY data taken at $R\approx4000$, $\mathrm{FOV}_{\mathrm{interf}} \sim 7"$ for $\lambda=2\text{ $\mu$m}$ and $B=130\text{ m}$ and this is a completely negligible effect.
\end{enumerate}

\subsection{How we find interferometric companions} 

\subsubsection{A simple binary} 

If the source is only a binary, the square visibilities (as a function of wavelength or spatial frequency) in each baseline are a sinusoidal that varies between 1 and $\left ( \frac{1-f}{1+f} \right )^2$, where $f$ is the flux ratio between the components. A binary nature can therefore be directly inferred from simple inspection of the interferometric data. 

However, because our interferometric data consists of squared visibilities and closure phases (i.e. we do not have the visibility phases necessary to assemble the complex visibility), the $\chi^2$ map produced by fitting Eqn. \ref{eqn:multiple} to the data is not convex and has multiple local minima. The usual approach to overcome this, which we also follow, is to construct 2d grids in the binary separation in order to find the approximate location of the global minimum, followed by a gradient-based fit to find the exact global minimum. This is very standard practice in fitting a binary model to optical/NIR interferometric data \citep[see e.g.][]{Gallene15}. We note that all our fits are performed using the \texttt{python} package \texttt{lmfit}.

\subsubsection{Triples and beyond} 

A 3+ multiplicity can be directly inferred from the data in case the sinusoidal components in the squared visibilities do not reach unity and are not exactly periodic. This is because each new component $i$ adds $i$ sinusoidal components with different oscillation frequencies to the model equation \ref{eqn:multiple}. This turned out to be the case for two of the targets in this paper (HIP 5300 and HIP 47479). 

If the additional component(s) are significantly fainter than the primary and secondary, they add a small perturbation which is not enough to displace the global minimum from the model where only the binary exists. In this case, one can find the global minimum for a binary model, fix the parameters for the primary and secondary, and run another grid to find the location of the global minimum for the tertiary. However, if the latter has a flux comparable to the other components, it shifts the global minimum enough (since the binary model will try to compensate for its presence) that this iterative approach fails, so that a larger dimensional grid for multiple components is needed. 

We note that one possible way to avoid this would be to estimate the visibility phases using the closure phases, followed by deconvolution-based imaging (e.g. \texttt{CLEAN} algorithm with phase self-calibration) to find the basic architecture of the system (number of components and their approximate location), to be followed by model-fitting for the precise parameters. This approach was used, for example, in \cite{Hummel03} for the triple A-star system $\eta$ Virginis. In our case, however, we found that this method did not work because of the relatively small number of telescopes (four, so that the closure phases only contain 50\% of the visibility phase information) as well as the rather short observations, which result in a poor $uv$ coverage and very nontrivial dirty beam. Therefore, we proceeded with the grid search approach followed by model fitting.

 
 \subsubsection{Faint companion detection limits}
 
 A single star will ideally have unity squared visibilities and zero closure phases for any $uv$ coordinate. In practice, the errors in the interferometric measurables are dominated by systematic errors due to imperfect calibration of the squared visibilities (the closure phases tend to be more robust) caused by the change in the transfer function with time. By comparing the data across the different files, we found that often the threshold for a clear companion detection are squared visibilities below about 98\% and closure phases with absolute value above about $1^{\degr}$.
 
In the cases in which no clear companion could be detected, we follow \cite{Absil11} to calculate flux ratio upper limits for a putative companion as a function of separation. Namely, we first re-scale the error bars so that $\chi^2_{\mathrm{red}} = 1$ for a single star model. This calibrates the error bars to take into account systematic errors (including correlations between close spectral channels). Next, we construct a 2D grid in $1 \text{ mas} < \rho < 320 \text{ mas}$ and $0^{\degr} < \text{PA} < 360^{\degr}$ and, for each point on the grid, inject an additional source in the model and find the flux ratio at which the model becomes inconsistent with the data at the $3\sigma$ level. Due to the sparse $uv$-coverage, the flux ratio upper limit at a given separation depends on the PA (in other words, there are specific positions where the source can be injected and lead to smaller interferometric signatures). Therefore, for each separation we find the median and the $90\%$ flux limits over the full range of PAs. In general, the contrast upper limit degrades at low separations due to the limited angular resolution of the array and at large separations due to the attenuation of the light from the source coupled into the GRAVITY fiber.

\subsection{CPM companion search with \textit{Gaia}} 

The VAST survey \cite{DeRosa14} includes a common proper motion (CPM) search for a limited set of the AO targets. Out of the 108 A stars in our parent sample, only 20 of them were also part of the CPM sample. Furthermore, out of the three CPM companions detected among these 20 targets, we later show that two of them are not real companions. 

We use \textit{Gaia} eDR3 in order to perform a CPM search for the 108 targets in our parent sample (except the few of them which have no \textit{Gaia} solution because they are known or suspected multiples). Given the distances $d \lesssim 75 \text{ pc}$, \textit{Gaia} should be complete down to extremely faint objects (including white dwarfs) for separations $\rho \gtrsim 5"$ and provides extremely precise parallaxes and proper motions. For each target, we select all \textit{Gaia} eDR3 sources which have projected separations within $10^6 \text{ AU}$ of the target and 3d distances less than 10 pc, and plot them on a proper motion difference (in $\text{ km} \text{ s}^{-1}$) versus projected separation (in AU) plane, colored by the significance in their parallax difference (estimated from the parallax errors). Owning to \textit{Gaia}'s extreme precision, we can then readily identify real CPM companions that fall near or below the mutual Keplerian velocity line defined by 

\begin{equation}
v_{\mathrm{kep}} = \left ( \frac{G \times 3 M_{\odot}}{ \frac{\sqrt{3}}{\sqrt{2}} \rho_{\mathrm{proj}} } \right )^{1/2} 
\end{equation}

\noindent where $3 M_{\odot}$ is a rough estimate for the total mass in the system and the square root factors account for the average correction factor from projected to real separation. 
In almost all cases, a companion that falls below or near the Keplerian line has a consistent parallax with the A star. In just one case, the parallax was formally inconsistent but the difference of a few parsecs is probably due to the companion having a large RUWE (and therefore possibly being a binary); in this case, we still considered it to be a CPM companion. 

Once a CPM companion is identified, we estimate its mass using its absolute \textit{Gaia} g band magnitude assuming it is a MS star and interpolating from the table made public by Erik Mamajek\footnote{\url{http://www.pas.rochester.edu/~emamajek/EEM_dwarf_UBVIJHK_colors_Teff.txt}} \citep{Pecaut12,Pecaut13}. We check that the \textit{Gaia} $B_p - R_p$ color is consistent with a MS star, which it is in all cases except for the one white dwarf companion within our sample. We also used the referred table for assigining a spectral type to the targets based on their inferred masses.

\section{Interferometry results on individual systems}

In this section, for each of the objects we observed with GRAVITY we briefly summarize its status prior to this paper and present our new companion detections or upper limits. 

In order to convert K band flux ratios into masses, we proceed in the following way. For stars with $2.5 \lesssim M_K \lesssim 6.5$ ($0.4 M_{\odot} \lesssim M \lesssim 1.5 M_{\odot}$), there exists a rather tight log-linear K band magnitude-mass relation \citep[e.g.][]{Henry93}, with small deviations for very young stars (age $\lesssim 100 \text{ Myrs}$). Using Figure A.5 (right) in \cite{DeRosa14}, which was constructed from theoretical isochrones from \cite{Baraffe98}, we find an approximate relation 

\begin{align}
\label{eqn:Kband_to_mass} 
\log M [M_{\odot}] \approx -0.145 M_{K} + 0.496  
\end{align}

\noindent which we use to estimate the mass of companions in this mass range. For more massive stars, the age effect is much more important and a precise mass cannot be obtained from the absolute K band magnitude alone. In this case, we use the K band absolute magnitude combined with the $M_{V_T}-M_K$ color to estimate both the mass and the age of the star following Figures A.3 and A.4 in \cite{DeRosa14}, which were constructed using theoretical isochrones with no rotation (so that the age estimates should be taken as upper limits). 

\begin{table*}
\centering
\caption{\label{table:fit_results} Best-fit parameters for the interferometric multiple star models.}
\begin{tabular}{ccccc}
\hline \hline
target & \shortstack{$f_i / f_1$} & \shortstack{$\Delta \alpha*_{i1}$\\(mas)} & \shortstack{$\Delta \delta_{i1}$\\(mas)} &  $\chi^2_{\mathrm{red}}$ \\[0.3cm]

HIP 5300 & \shortstack{$f_{21}=0.150\pm0.002$\\$f_{31}=0.234\pm0.001$\\$f_{41}=0.188\pm0.001$} & \shortstack{$\Delta \alpha*_{21}=6.527\pm0.006$\\$\Delta \alpha*_{31}=-231.74\pm0.01$\\$\Delta \alpha*_{41}=-233.89\pm0.02$} & \shortstack{$\Delta \delta_{21}=-3.87\pm0.01$\\$\Delta \delta_{31}=194.90\pm0.01$\\$\Delta \delta_{41}=193.84\pm0.02$} & $7.5$ \\[0.3cm]

HIP 47479 & \shortstack{$f_{21}=0.108\pm0.001$\\$f_{31}=0.703\pm0.003$\\$f_{41}=0.248\pm0.003$} & \shortstack{$\Delta \alpha*_{21}=3.68\pm0.03$\\$\Delta \alpha*_{31}=-114.00\pm0.01$\\$\Delta \alpha*_{41}=-116.66\pm0.03$} & \shortstack{$\Delta \delta_{21}=-1.52\pm0.01$\\$\Delta \delta_{31}=43.536\pm0.004$\\$\Delta \delta_{41}=43.344\pm0.009$} & $2.4$ \\[0.3cm]

HIP 54477 & $0.0380\pm0.0002$ & $-140.181\pm0.009$ & $100.905\pm0.009$ & $3.0$ \\[0.3cm]

HIP 54746 & $0.02043\pm0.00004$ & $13.933\pm0.002$ & $-11.045\pm0.003$ & $3.1$ \\[0.3cm]

HIP 70931 & $0.948\pm0.001$ & $-1.2331\pm0.0003$ & $-0.9572\pm0.0008$ & $6.3$ \\[0.3cm]

HIP 104139 & $0.01180\pm0.00004$ & $-15.136\pm0.008$ & $7.740\pm0.009$ & $4.9$ \\[0.3cm]

HIP 116758 & $0.09464\pm0.00008$ & $20.801\pm0.002$ & $-15.906\pm0.004$ & $3.6$ \\[0.3cm]

HIP 118092 & $0.0658\pm0.0001$ & $118.724\pm0.005$ & $39.099\pm0.007$ & $2.6$ \\[0.3cm]

\hline
\end{tabular}
\end{table*}

\subsection{HIP 5300 ($\upsilon$ Phe)} 

\cite{McDonald12} found $T_{\mathrm{eff}}=8000 \text{ K}$ and $L=21.0 L_{\odot}$ ($R=2.4 R_{\odot}$). It has a spectral type A4V in \cite{Abt95}. \cite{McDonald12} also found an infrared ($\lambda > 2.2 \mu$m) excess by a factor of 1.7, which could point to a circumstellar disk or additional companion. However, this is a known AO binary \citep[WDS 01078-4129; ][]{Soderhjelm99,Horch01,DeRosa14} so these results are biased. 

Based on the contrasts $\Delta V=1.34$ and $\Delta K=0.57$, \cite{DeRosa14} estimated masses $M_1=2.10 M_{\odot}$ and $M_2=1.64 M_{\odot}$ and age $t=400 \text{ Myrs}$. They also re-derived the orbital parameters using only high spatial resolution (AO and speckle interferometry) observations. The period $P=28.36 \pm 0.04 \text{ yrs}$ and semi-major axis $a=0.2396" \pm 0.0005$ (16.8 AU) imply a large total dynamical mass $M_{\mathrm{dyn}}=6.0 \pm 1.0 M_{\odot}$ when combined with the distance in Table \ref{table:targets}. The discrepancy with the spectroscopic masses is a very strong hint that one or both of the companions are themselves multiples.

\begin{figure}
 \includegraphics[width=\columnwidth]{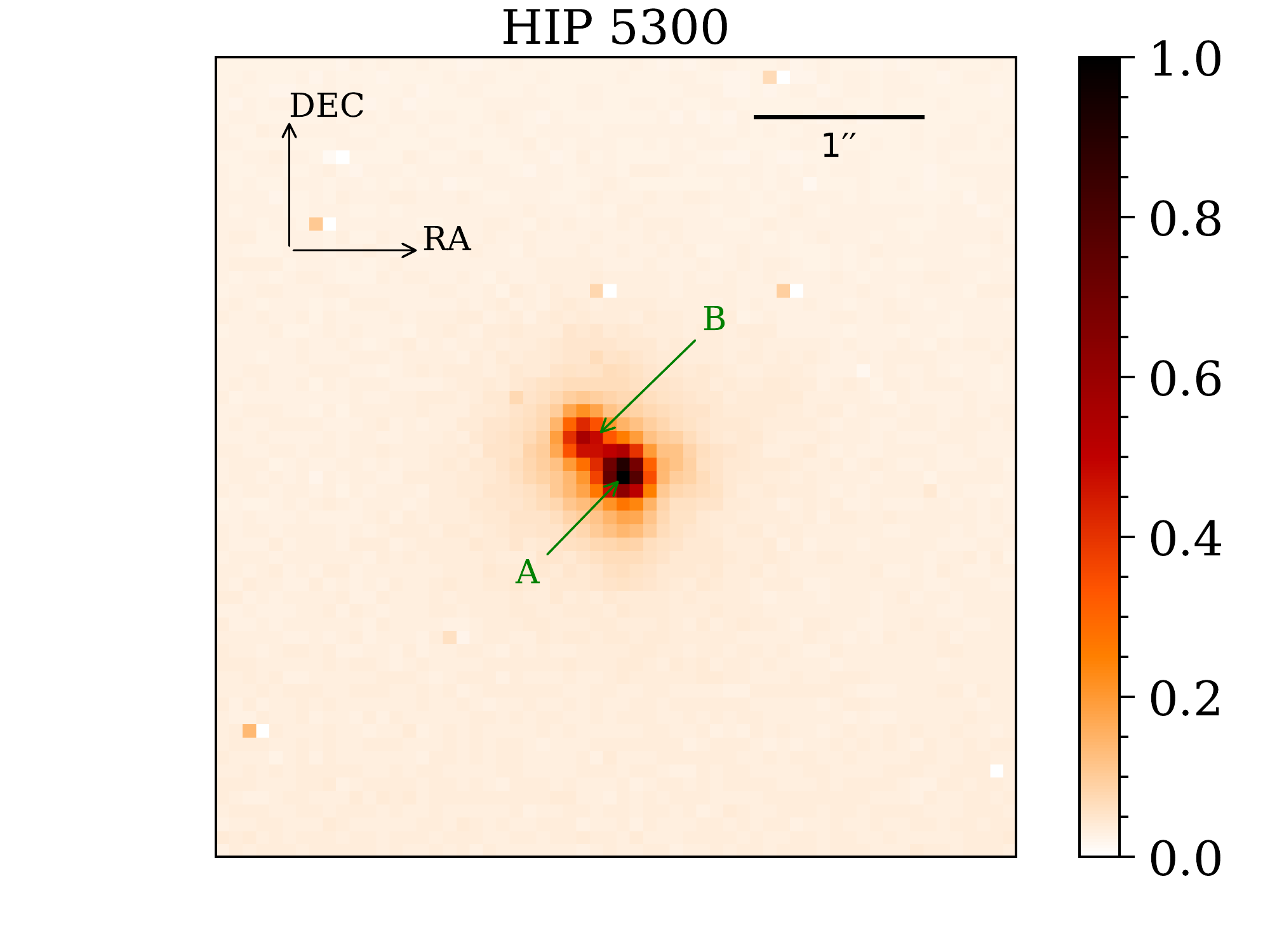}
 \caption{GRAVITY acquisition camera image of HIP 5300 showing the resolved and previously known AO binary.}
 \label{fig:acqcam_5300}
\end{figure}

Indeed, the interferometric data immediately shows that this must be at least a triple system. The very high frequency visibility oscillations are due to the known AO binary (which can actually also be seen in the AO image from the GRAVITY acquisition camera, shown in Figure \ref{fig:acqcam_5300}), but there are clear lower frequency modulations caused by additional closer companion(s) that prevent the squared visibilities from reaching unity. 

\begin{figure*}
 \includegraphics[width=2.2\columnwidth]{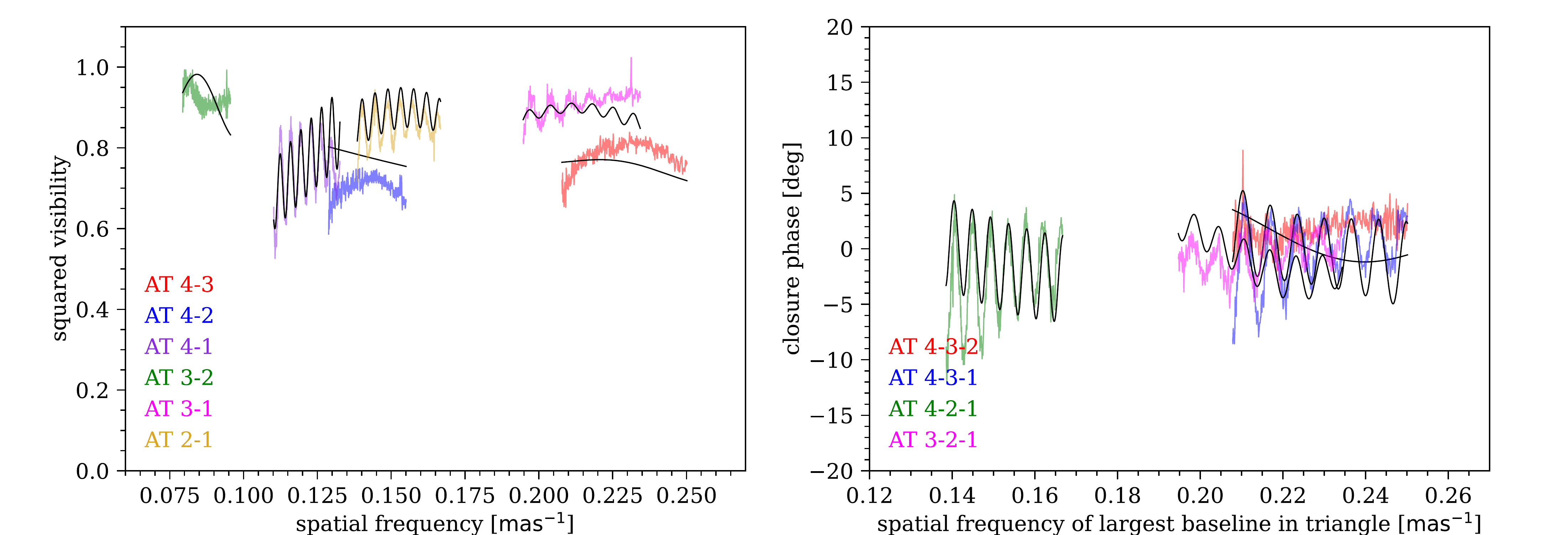}
 \caption{Inteferometric data (colored) and best-fit 2+2 quadruple model (black) for HIP 5300.}
 \label{fig:model_fit_5300}
\end{figure*}

A binary fit recovers the AO binary but is a very poor fit to the data ($\chi^2_{\mathrm{red}}=19.8$). Using a triple system model where either the primary or the secondary are allowed to be close binaries, we find an improvement to $\chi^2_{\mathrm{red}}=12.6$ and $\chi^2_{\mathrm{red}}=10.9$, respectively. Using a quadruple 2+2 model, where both the primary and secondary are close binaries, we arrive at $\chi^2_{\mathrm{red}}=7.5$. This model can explain most albeit not all of the features in the interferometric data (Fig. \ref{fig:model_fit_5300}). We suspect there might be additional complexity needed for modeling this system because the separation of the AO binary is larger than the FWHM of the GRAVITY fiber; therefore, the amount of flux of the secondary coupled into the fiber can be very sensitive and might be different for different telescopes, making the flux ratios telescope dependent. Although this can in principle be modelled \citep[e.g.][]{Waisberg19}, in the case of a multiple system such as HIP 5300 this would make the model parameters very degenerate and is beyond the scope of this paper. Therefore, we consider our best fit solution quite preliminary and further observations are needed to confirm it. Ideally, the Unit Telescopes would be used so that each component in the AO binary could be observed separately and the degeneracy and complexity would then be easily broken. The fact that this target is at least a triple system, however, is a robust conclusion from the current data. 

The best-fit parameters for the 2+2 quadruple solution are listed in Table \ref{table:fit_results}. The K band flux ratio between the two binaries is 0.37, compared to 0.59 reported in \cite{DeRosa12}. The large discrepancy is probably caused by a large systematic error in the flux ratio due to the large separation compared to the FWHM of the GRAVITY fiber. The position of the centroid of the fainter binary relative to that of the brighter one is $\rho = 304.2 \text{ mas}$ and $\mathrm{PA} = 309.8^{\degr}$, compared to a predicted $\rho = 305.9 \text{ mas}$ and $\mathrm{PA} = 309.6^{\degr}$ based on the orbital parameters in \cite{DeRosa12}. We note that by performining PSF deblending on the acquisition camera image shown in Fig. \ref{fig:acqcam_5300}, we obtain a K band flux ratio of 0.47 and a separation of $\rho = 318 \text{ mas}$ and $\mathrm{PA} = 311^{\degr}$ for the AO binary. Clearly, further observations are needed to resolve the small inconsistencies. 

Based on our very preliminary interferometric model, the primary is a close binary with K band flux ratio of 0.15 and projected separation of 4 mas (0.53 AU), while the secondary is another close binary with K band flux ratio 0.80 and projected separation 2.4 mas (0.17 AU). Using the more robust flux ratio between the secondary and primary binaries of 0.59 from the AO observations in \cite{DeRosa14} and the relative flux ratios in each binary from the interferometric fit, we find K band absolute magnitudes of 1.20 and 3.26 for the primary binary and 2.25 and 2.49 for the secondary binary. For the three fainter components, we can use relation \ref{eqn:Kband_to_mass} to estimate masses of about $1.05 M_{\odot}$, $1.36 M_{\odot}$ and $1.48 M_{\odot}$. Furthermore, using the absolute $V_T=1.28$ magnitude of the primary binary, and using the approximation that the optical light is dominated by the A star, we find a mass of around $2.2 M_{\odot}$ and an age $t \sim 320 \text{ Myrs}$ for the brightest star in the system. Therefore, the system would be composed of A0+G0 and F1+F4 binaries, with a total mass of $6.1 M_{\odot}$, in agreement with the dynamical mass inferred from the outer orbit. 

\subsection{HIP 47479}

\cite{Zorec12} found $T_{\mathrm{eff}}=7980\pm239 \text{ K}$, $L=37.5 \pm 2.5 L_{\odot}$ ($R=3.2 R_{\odot}$), $M=2.23\pm0.03 M_{\odot}$ and fractional age on the MS $t_{\mathrm{MS}}=0.855\pm0.049$. It has spectral type A3IV and projected rotational velocity $v \sin i = 10 \text{ km}\text{ s}^{-1}$ \citep{Royer07}. However, this is a known close AO binary \citep[WDS 09407-5759; ][]{Soderhjelm99,Mason01,DeRosa14} and therefore these values are biased. Based on the contrasts $\Delta V = 0.60$ and $\Delta K = 0.08$ for the close binary, \cite{DeRosa14} estimated masses $M_1=2.18$ and $M_2=2.10 M_{\odot}$ and age $t=560 \text{ Myrs}$. They also re-derived the orbital parameters using only the more reliable high spatial resolution (AO and speckle interferometry) observations. The period $P=10.74 \pm 0.04 \text{ yrs}$ and semi-major axis $a=0.1207" \pm 0.0006"$ (8.2 AU) imply a total dynamical mass $M_{\mathrm{tot}}=4.8\pm0.4 M_{\odot}$ using the distance in Table \ref{table:targets}.

\begin{figure*}
 \includegraphics[width=2.2\columnwidth]{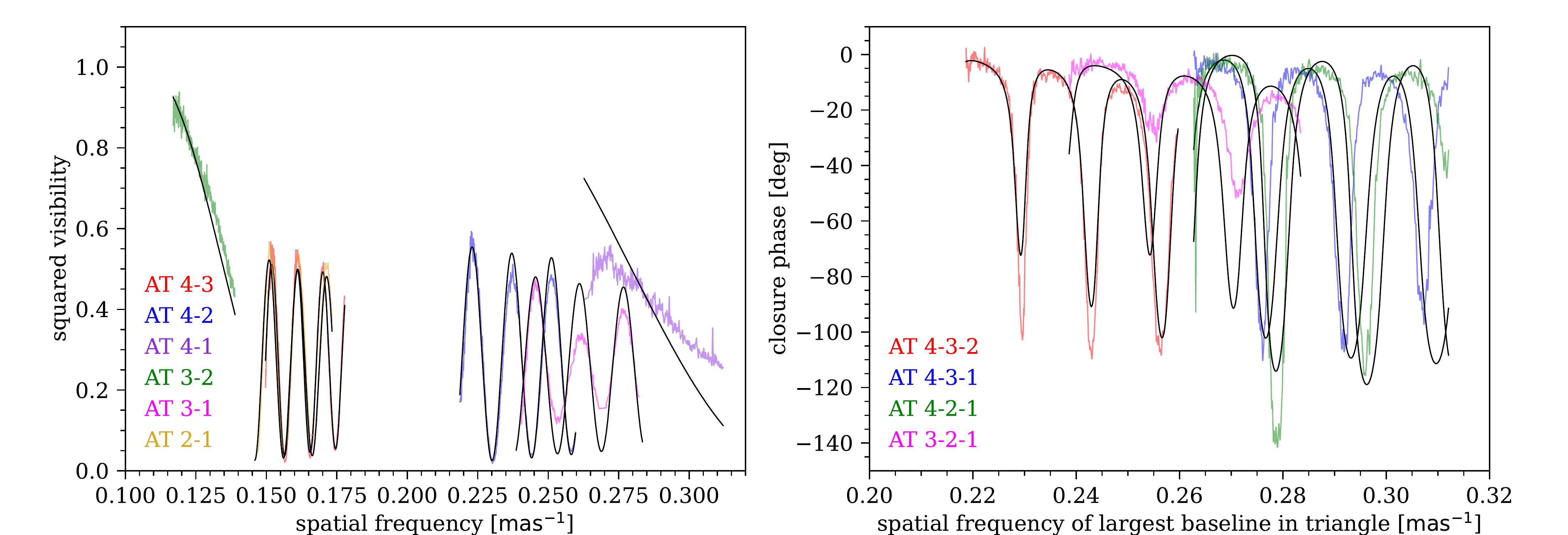}
 \caption{Inteferometric data (colored) and best-fit 2+2 quadruple model (black) for HIP 47479.}
 \label{fig:model_fit_47479}
\end{figure*}

The interferometric data immediately reveals that a binary model is inadequate ($\chi^2_{\mathrm{red}}=45.5$): the high frequency oscillations in the visibility are caused by the AO binary at large separation, but again an additional companion(s) is(are) needed to prevent the squared visibilities from reaching unity. Using triple system models in which the primary or the secondary are close binaries, we find an improvement to $\chi^2_{\mathrm{red}}=6.6$ and $\chi^2_{\mathrm{red}}=4.3$, respectively. Finally, allowing both the primary and secondary stars to be close binaries in a quadruple system model, we find an improvement to $\chi^2_{\mathrm{red}}=2.4$, which is formally significant enough to be preferable over the triple star models. Our final model for HIP 47479 is thus a 2+2 quadruple system and is shown together with the interferometric data in Figure \ref{fig:model_fit_47479}. The model explains most of the features of the data, but still leaves some unexplained systematic structure, especially in the closure phases. This might be related to even more components or to a mild contribution from an extended disk to the K band light, given that a circumstellar or circumbinary disk exists in this system from its significant mid-infrared excess at $25$ and $60 \mu$m \citep{Cheng92}. Clearly, further observations are needed to more accurately characterize this very complex system. 

The best-fit parameters for our 2+2 quadruple solution are listed in Table \ref{table:fit_results}. The K band flux ratio between the two close binaries is $0.86$ (compared to 0.93 reported in \cite{DeRosa12}) and the position of the fainter binary relative to the brighter one is $\Delta \rho = 123.1 \text{ mas}$ and $\mathrm{PA} = 290.9^{\degr}$, compared to a predicted $\Delta \rho = 117.5 \text{ mas}$ and $\mathrm{PA} = 291.8^{\degr}$ based on the orbital parameters in \cite{DeRosa12}. The primary is a close binary with K band flux ratio of 0.11 and projected separation of 4 mas (0.27 AU), while the secondary is another close binary with K band flux ratio 0.35 and projected separation 2.8 mas (0.19 AU). With these flux ratios, we find absolute K band magnitudes of 1.42 and 3.82 for the stars in the primary binary and 1.81 and 2.92 for those in the secondary binary. For the two fainter components in each binary, we can use relation \ref{eqn:Kband_to_mass} to estimate masses of about $0.9 M_{\odot}$ and $1.2 M_{\odot}$, respectively. Furthermore, using the absolute $V_T$ band magnitude in each binary, 1.67 and 2.26, and using the approximation that the optical light is dominated by the A stars, we find masses of around $2.1 M_{\odot}$ and $1.8 M_{\odot}$ for the brighter star in the primary and secondary binaries, respectively, and an age $t \sim 250 \text{ Myrs}$. According to the model this system is therefore composed of A1+G9 and A7+F7 binaries. The total mass of around $5.9 M_{\odot}$ is at 3$\sigma$ tension with the dynamical mass from the outer orbit, which is another motivation for further observations of this system. 

Finally, we note that \citep{Shaya11} reported a candidate very wide companion for HIP 47479, a K3III star (HIP 46701) at separation 6158" (419000 AU), with a 67\% association probability based on common parallaxes and proper motion. Based on our CPM search using \textit{Gaia} eDR3, including subtracting the proper motion of the center of light of HIP 47479 at the \textit{Gaia} eDR3 epoch due to its 10.74 yrs orbit and using the mass ratio from the model fit above and the V band flux ratio as an estimate for the \textit{Gaia} g band flux ratio, we find that the relative proper motion of this companion is around two orders of magnitude above the Keplerian value, and therefore we exclude it as a physical companion. Moreover, no other CPM candidate is found for HIP 47479. 

\subsection{HIP 54477 (10 Crt)} 

\cite{Zorec12} found $T_{\mathrm{eff}}=8954\pm144 \text{ K}$, $L=20.7 \pm 1.4 L_{\odot}$ ($R=1.9 R_{\odot}$), $M=2.06\pm0.03 M_{\odot}$ and fractional age on the MS $t_{\mathrm{MS}}=0.315\pm0.075$. \cite{DeRosa14} reported similar values of $M=2.09 M_{\odot}$ and age $t=250 \text{ Myrs}$. It has spectral type A1V and projected rotational velocity $v \sin i = 249 \text{ km}\text{ s}^{-1}$ \citep{Royer07}. This star had no known companion prior to this paper. 

\begin{figure*}
 \includegraphics[width=2.2\columnwidth]{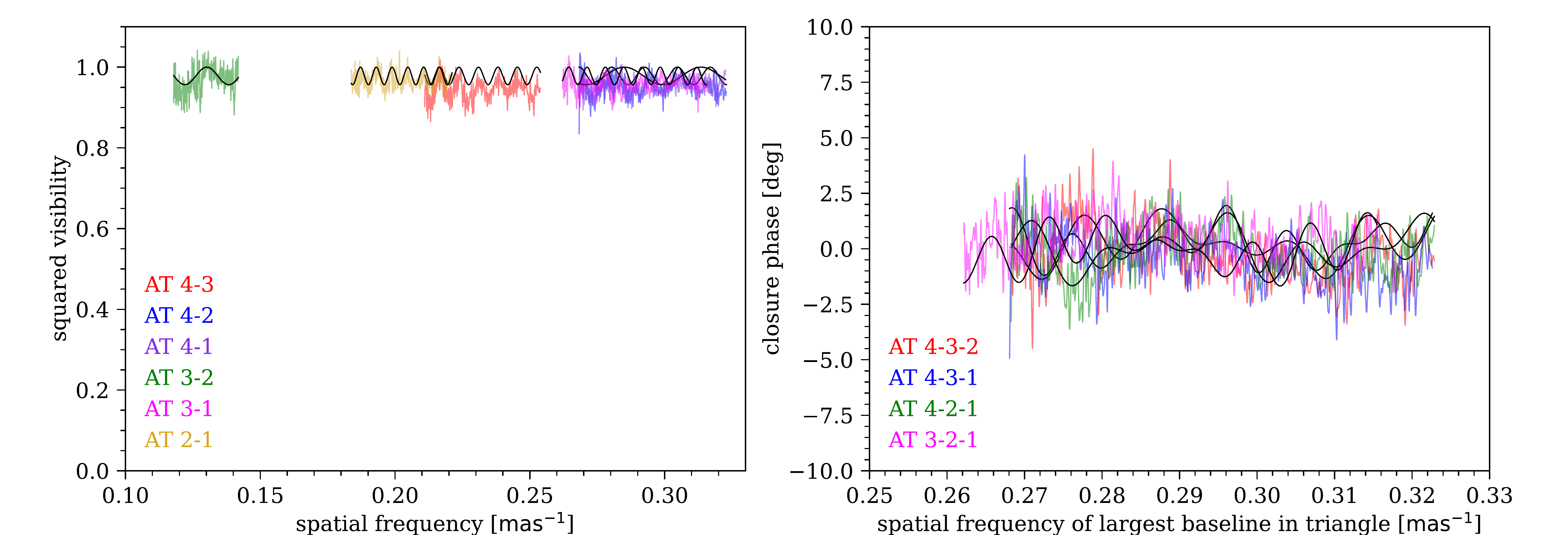}
 \caption{Inteferometric data (colored) and best-fit binary model (black) for HIP 54477.}
 \label{fig:model_fit_54477}
\end{figure*}

This star is located near the upper branch in Fig. \ref{fig:Hip-Gaia}, so we expect a period of a few decades. The interferometric data shown in Fig. \ref{fig:model_fit_54477} reveals the presence of a faint companion with a large projected separation causing a high frequency oscillation in the squared visibilities. The best-fit binary model, with parameters reported in Table \ref{table:fit_results}, have a K band flux ratio $0.038$ and projected separation $\rho = 172.7 \text{ mas} \leftrightarrow 9.4 \text{ AU}$. 

The absolute K band magnitude of the companion is 5.14 and translates to a mass of $0.56 M_{\odot}$ from Eq. \ref{eqn:Kband_to_mass}. The companion is faint enough to have a negligible bias in the reported properties of the A star. This is therefore an A1V+M0V binary. The projected separation and the total mass of 2.65 $M_{\odot}$ translate to a period estimate of 17.7 yrs from Kepler's Third Law.  

\subsection{HIP 54746} 

This star is classified in SIMBAD as A3IV-V. \cite{McDonald12} found $T_{\mathrm{eff}}=8103 \text{ K}$ and $L=14.12 L_{\odot}$ ($R=1.9 R_{\odot}$). \cite{DeRosa14} reports a mass $M=1.96 M_{\odot}$ and age $t=400 \text{ Myrs}$. Its rotational velocity is $v \sin i = 146 \text{ km}\text{ s}^{-1}$ \citep{Glebocki05}. 

This star is located near the lower branch in Fig. \ref{fig:Hip-Gaia}, so we expect it to be a close binary with a period of around 1 year. Indeed, this star is a known astrometric binary (WDS 11126-4906) from \textit{Hipparcos} with period $975_{-44}^{+58} \text{ days}$ and centroid semi-major axis $14_{-3}^{+23} \text{ mas}$ \citep{Goldin06}; this close companion appears to have been missed in \cite{DeRosa14} literature search. It is also an X-ray source with $L_X \sim 2 \times 10^{29} \text{ erg} \text{ s}^{-1}$ \citep{Schroder07}, which suggests the presence of a low mass companion. We note that Table 9 in \cite{DeRosa14} refers to a 3.80" companion to HIP 54746; however, we found no such information in the reference provided and there is no evidence for such companion either in the GRAVITY acquisition camera image or in our \textit{Gaia} common proper motion search. Therefore, we suspect their reporting was a mistake. 

\begin{figure*}
 \includegraphics[width=2.2\columnwidth]{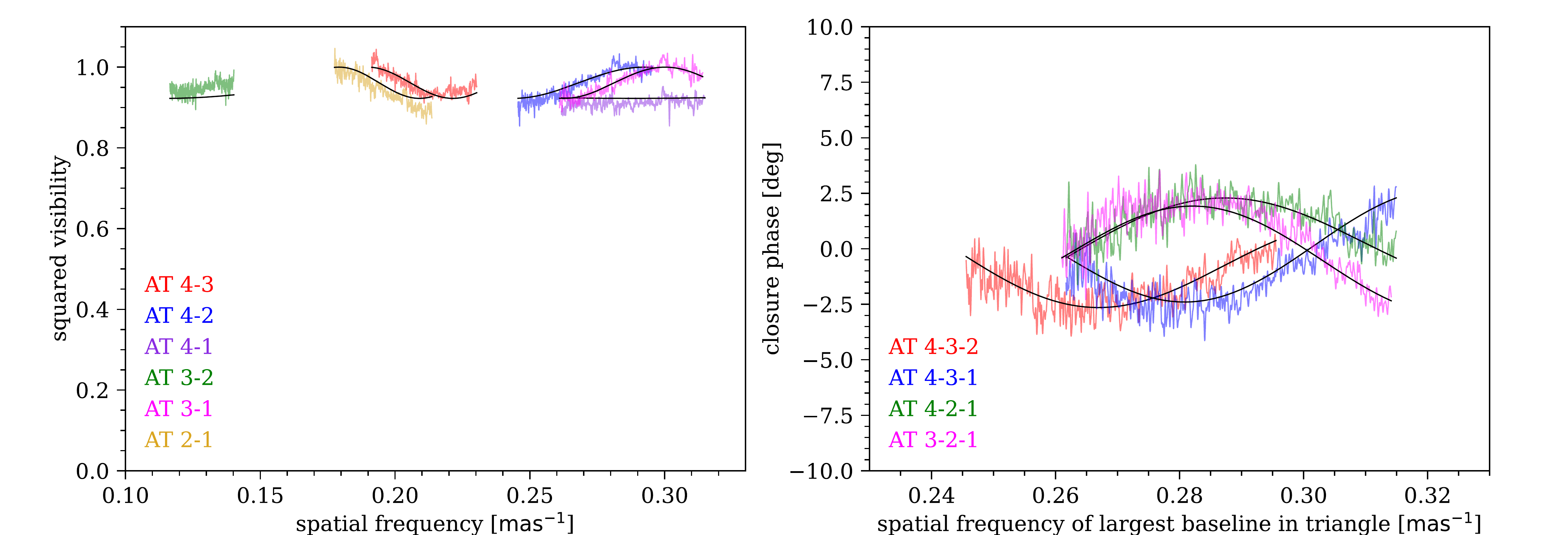}
 \caption{Inteferometric data (colored) and best-fit binary model (black) for HIP 54746.}
 \label{fig:model_fit_54746}
\end{figure*}

The interferometric data, shown in Fig. \ref{fig:model_fit_54746} along with the best-fit binary model, indeed reveals a faint companion with a K band flux ratio $0.020$ and projected separation $\rho = 17.8 \text{ mas} \leftrightarrow 0.97 \text{ AU}$. The absolute K band magnitude of the companion is 5.54 and translates to a mass of $0.49 M_{\odot}$ from Eq. \ref{eqn:Kband_to_mass}. The companion is faint enough to have a negligible bias in the reported properties of the A star. This is therefore an A3V+M1V binary. The projected separation and the total mass of 2.45 $M_{\odot}$ translate to a period estimate of 223 days from Kepler's Third Law, which is about a factor of four smaller than the period inferred from \textit{Hipparcos} astrometry but this is not a surprise since it is a high inclination system $i = 68^{\degr} \pm 10^{\degr}$ \citep{Goldin06}. Therefore, we have resolved the \textit{Hipparcos} astrometric binary. 

We note that this star was detected as an X-ray source in the ROSAT all-sky survey \citep{Voges00} with a positional error of 14". The count rate of $0.078 \pm 0.024 \text{ photons} \text{ s}^{-1}$ within $0.1-2.4$ keV was converted to an X-ray luminosity $L_X \sim 2 \times 10^{29} \text{ erg} \text{ s}^{-1}$ in the catalogue of X-ray emitting A stars of \cite{Schroder07}. The X-ray emission can be explained by the M1V companion with a rotation period $\lesssim 10 \text{ days}$ \citep{Magaudda20}, which is expected since the A star requires a young age for the system. 

\subsection{HIP 70931}

\cite{McDonald12} reported $T_{\mathrm{eff}}=8560 \text{ K}$ and $L=21.94 L_{\odot}$ ($R=2.1 R_{\odot}$) for this A star. \cite{DeRosa14} reported the discovery of a faint AO companion with contrast $\Delta K = 4.63$ and separation $\rho = 0.60"$, estimating masses $M_1=1.77 M_{\odot}$ and $M_2=0.23 M_{\odot}$ and a very young age $t=40 \text{ Myrs}$ for the system. However, the primary is a known close double-lined spectroscopic binary with a period 11.8 days, mildly eccentric (e=0.3) and with a mass ratio q=0.93 \citep{Kaufmann73}. The masses are estimated to be $1.84 M_{\odot}$ and $1.71 M_{\odot}$ (A6V+A9V) in the Multiple Star Catalog \citep{Tokovinin18} based on the absolute V magnitude and the spectroscopic mass ratio, implying an orbit close to edge on ($\sin i \approx 0.98$) and a semi-major axis $a=0.15 \text{ AU}$. The projected rotational velocities of the A stars are claimed to be small, $v \sin i \lesssim 15 \text{ km}\text{ s}^{-1}$ \citep{Kaufmann73}, which is interesting as it suggests synchronization even though $\frac{a}{R}=16$ appears too high to allow tidal synchronization on such a young system with radiative envelopes \citep[e.g.][]{Zahn77}. We note that the very faint AO companion reported by \cite{DeRosa14} is not visible in the GRAVITY acquisition camera image due to the 5-magnitude attenuation filter used (to avoid saturation by the bright A star) and the short integration time of 1s. 

\begin{figure*}
 \includegraphics[width=2.2\columnwidth]{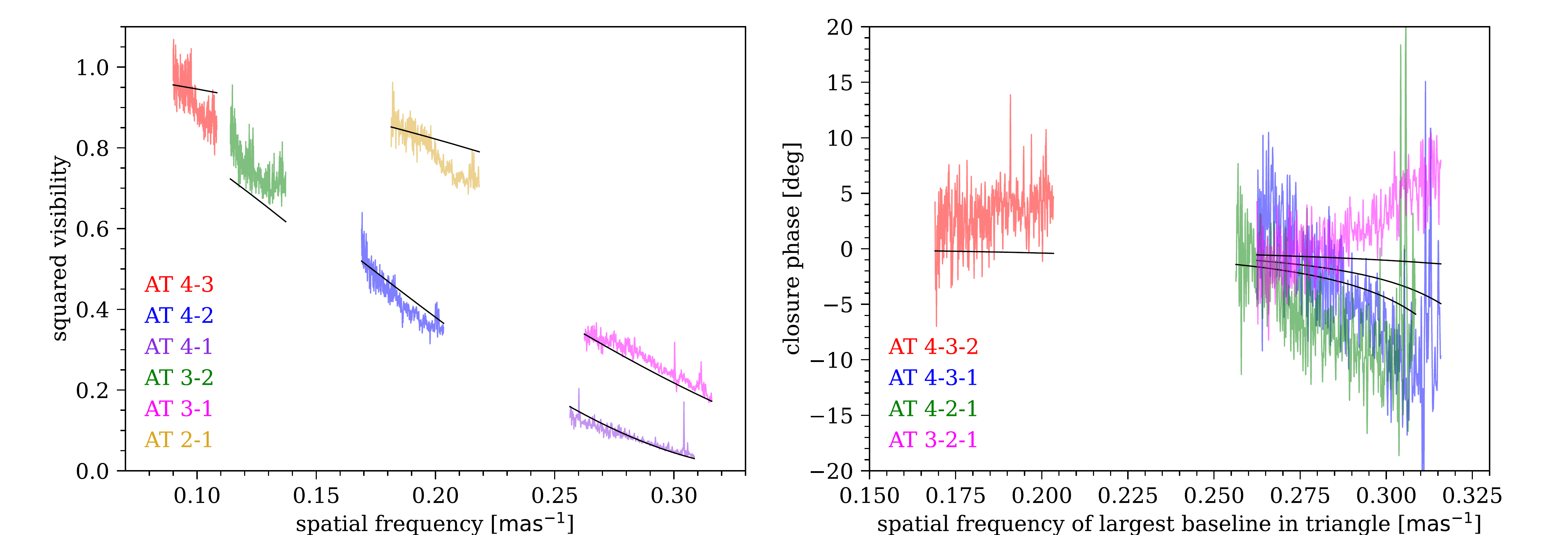}
 \caption{Inteferometric data (colored) and best-fit binary model (black) for HIP 70931.}
 \label{fig:model_fit_70931}
\end{figure*}

HIP 70931 is located near the lower branch in Fig. \ref{fig:Hip-Gaia}, with a large proper motion difference $\Delta v \gtrsim 1 \text{ km} \text{ s}^{-1}$ between \textit{Gaia} DR2 and eDR3. This suggests the presence of another unknown companion in the system with a period on the order of one year, since neither the faint AO companion (with an orbital period of many decades) nor the spectroscopic binary with a period of 11.8 days can in principle explain such a change. The interferometric data, shown in Fig. \ref{fig:model_fit_70931} along with the best fit binary model, is dominated by a very close binary with a K band flux ratio 0.95 and a projected separation $\rho = 1.56 \text{ mas} \leftrightarrow 0.095 \text{ AU}$, which can be readily identified as the known spectroscopic binary. We note that although the fit is not fully satisfactory ($\chi^2_{\mathrm{red}} = 6.3$), an extensive search for an additional companion with a triple star model did not result in any convincing detection. In particular, we notice that the closure phase residuals are similar in shape to the calibrator data, so that we believe the larger than expected residuals are due to calibration systematic errors (perhaps because the target is very resolved on two of the baselines due to the flux ratio near unity). 

We note that HIP 70931 was detected as an X-ray source with a positional error of 0.6" by ROSAT in the catalogue of pointed observations with the High Resolution Imager \citep[HRI;][]{ROSATHRI00}. The count rate of $0.0196 \pm 0.0010 \text{ photons} \text{ s}^{-1}$ within $0.1-2.4$ keV was converted to a large X-ray luminosity $L_X \sim 6.9 \times 10^{29} \text{ erg} \text{ s}^{-1}$ in the catalogue of X-ray emitting A stars of \cite{Schroder07}. By comparing the reported count rate and X-ray luminosity to other stars at various distances, we noticed that the reported X-ray luminosity for HIP 70931 is a factor of 10 higher than what it should be (most likely due to a typo), so that we assume that its actual X-ray luminosity is $L_X \sim 7 \times 10^{28} \text{ erg} \text{ s}^{-1}$. This is yet another evidence for a fourth body in this system, because the faint AO companion of $0.23 M_{\odot}$ does not appear to be massive enough to produce such X-ray luminosity even if it is fast rotating \citep[][]{Magaudda20}. We speculate that the upcoming \textit{Gaia} DR3 release might be able to reveal an unknown companion on a orbit of about 1 year if it exists. 

\subsection{HIP 77660 (b Ser)} 

\cite{Zorec12} found $T_{\mathrm{eff}}=8933\pm185 \text{ K}$, $L=26.2 \pm 1.3 L_{\odot}$ ($R=2.1 R_{\odot}$), $M=2.14\pm0.02 M_{\odot}$ and fractional age on the MS $t_{\mathrm{MS}}=0.479\pm0.074$ for this A star. It has spectral type A3V and projected rotational velocity $v \sin i = 229 \text{ km}\text{ s}^{-1}$. It is a known close visual binary with an orbital period of several decades \citep[WDS 15513-0305; e.g.][]{Mason10}, but the orbit is only partially covered, so that its parameters (in particular the period and therefore the dynamical mass) are very uncertain \citep{DeRosa12}. Based on the contrasts $\Delta V_T = 2.61$ and $\Delta K = 1.49$, \cite{DeRosa14} reports masses $M_1 = 2.05 M_{\odot}$ and $M_2 = 1.15 M_{\odot}$ and a system age $t = 350 \text{ Myrs}$. In addition, there have been reports of RV variability of HIP 77660 \citep{Iliev01,Nikolov08}. In particular, the latter suggest an orbit with $P = 1073 \text{ days}$, $e=0.7$ and $K_1 = 6 \text{ km} \text{ s}^{-1}$, with a resulting mass function $\frac{M_2^3}{(M_2+M_1)^2} \sin^3 i = 0.00875 M_{\odot}$. For the A star mass $M_1 = 2.05 M_{\odot}$, the resulting companion mass would then be $M_2 \gtrsim 0.37 M_{\odot}$. 

\begin{figure}
 \includegraphics[width=\columnwidth]{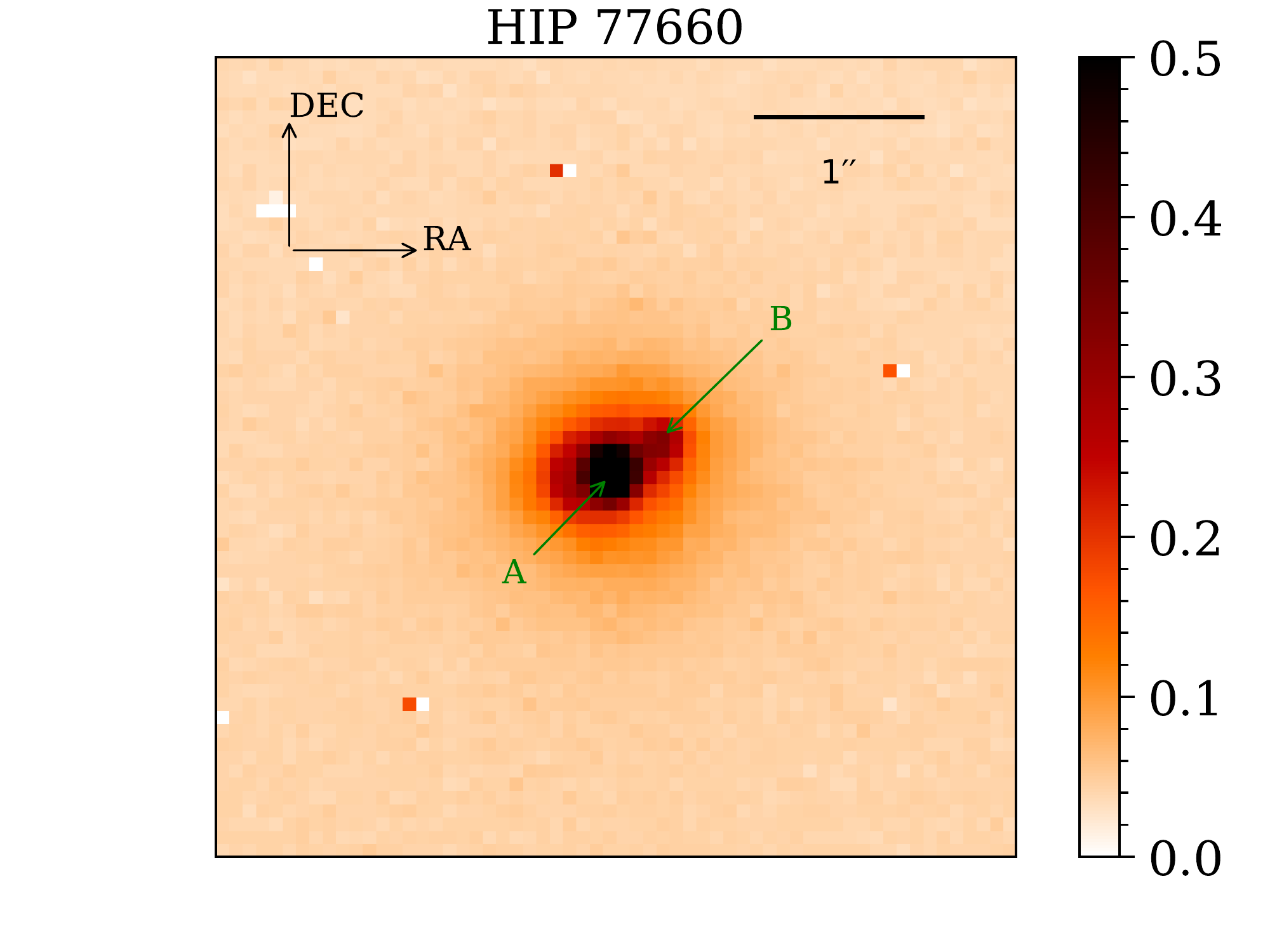}
 \caption{GRAVITY acquisition camera image of HIP 77660 showing the resolved and previously known AO binary.}
 \label{fig:acqcam_77660}
\end{figure}

The GRAVITY acquisition camera image (Figure \ref{fig:acqcam_77660}) shows the previously known AO companion at a separation $\Delta \alpha* = 365.2 \pm 2.1 \text{ mas}$ and $\Delta \delta = 205.6 \pm 3.1 \text{ mas}$ relative to the A star (found through PSF deblending). We checked that this position is consistent with that expected from the orbit and therefore it confirms that the objects are bound. At such a large separation, we expect that only around 0.04\% of the flux from the AO companion would be coupled into the GRAVITY fiber, resulting in a flux ratio of around 0.01\%, which is around two orders of magnitude below our detection limit, and therefore the effect of the AO companion on the interferometric data should be negligible. 

\begin{figure*}
 \includegraphics[width=2.2\columnwidth]{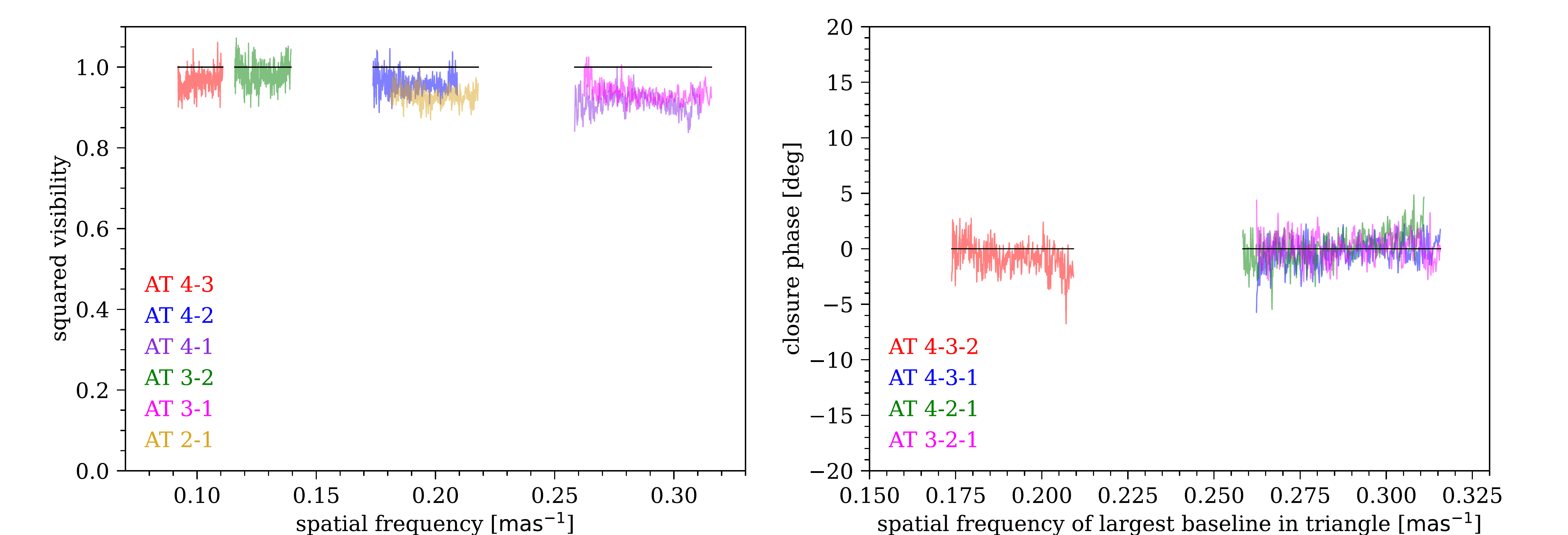}
 \caption{Inteferometric data (colored) for HIP 77660. The black lines show the expected signatures for a single star model.}
 \label{fig:model_fit_77660}
\end{figure*}

HIP 77660 is located near the lower branch in Fig. \ref{fig:Hip-Gaia}, with a large proper motion difference $\Delta v \gtrsim 2 \text{ km} \text{ s}^{-1}$ between \textit{Gaia} DR2 and eDR3, which is yet another evidence for a companion with an orbital period on the order of 1 year. The interferometric data, however, did not show any evidence for an additional companion. Fig. \ref{fig:model_fit_77660} shows the data and the black lines the expected quantities for a single star model. There is some loss of coherence for the longest baselines but a search for a binary companion revealed no solution that performs better than the single star model. 

\begin{figure}
 \includegraphics[width=\columnwidth]{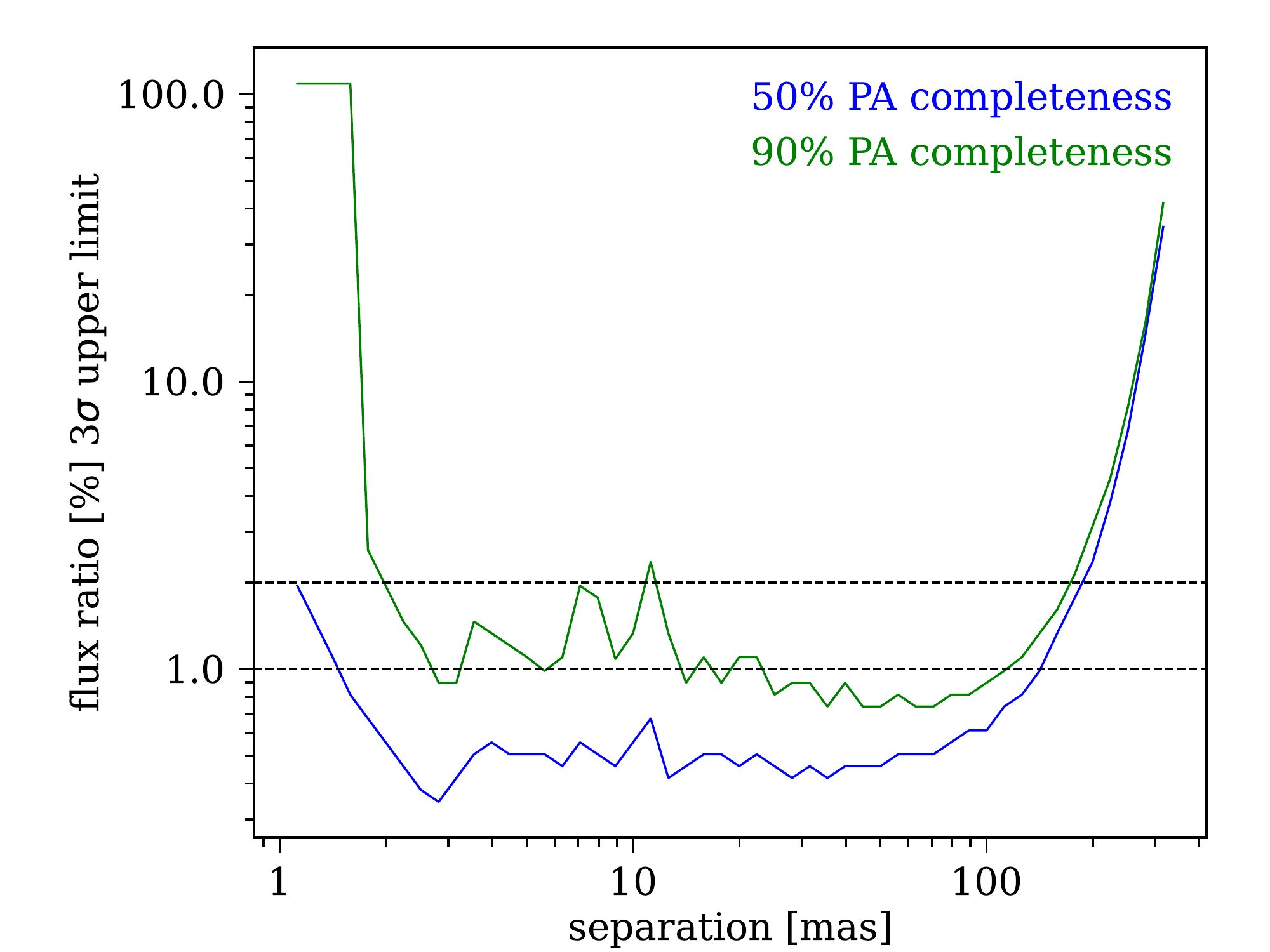}
 \caption{Flux ratio upper limit (3$\sigma$) for a companion in HIP 77660 based on the closure phases.}
 \label{fig:flux_limit_77660}
\end{figure}

Therefore, we proceed to find upper limits for the flux ratio of a possible companion, in this case using the closure phases only since there is a slight decrease in squared visibility at large spatial frequencies due to calibration systematic errors. The corresponding result is shown in Figure \ref{fig:flux_limit_77660}. We can exclude the presence of a companion with flux ratio above about 2\% with a projected separation $2 \text{ mas} \lesssim \rho \lesssim 150 \text{ mas}$. This corresponds to a lower limit on its absolute K band magnitude of 5.67, or a lower limit on its mass $M_2 \lesssim 0.47 M_{\odot}$ from Eqn. \ref{eqn:Kband_to_mass}. This is compatible with the mass function referred above as long as the orbital inclination is not too low. We speculate that the upcoming \textit{Gaia} DR3 release may be able to solve for this orbit.  

Finally, we note that HIP 77660 was detected in pointed observations with the PSPCs aboard ROSAT \citep{ROSAT00}. The count rate of $0.0419 \text{ photons} \text{ s}^{-1}$ within $0.1-2.4$ keV was converted to an X-ray luminosity $L_X = 5 \times 10^{28} \text{ erg} \text{ s}^{-1}$ in the catalogue of X-ray emitting A stars of \cite{Schroder07}. Given the young age of the system, the X-ray emission can be produced both by the known $1.15 M_{\odot}$ AO companion \citep{Garces11} and by the suspected lower mass spectroscopic companion \citep{Magaudda20}. 

\subsection{HIP 80628 ($\upsilon$ Oph)} 

The observations and analysis for this system will be presented in a separate dedicated paper (Waisberg et al. in prep) since they allow for interesting exploration of multi-body dynamics. Here, we briefly summarize the system since we include it in the multiplicity statistics of this paper. 

This is a known close visual binary \citep[WDS 16278-0822; ][]{Mason01,DeRosa14} with semi-major axis of 31.5 AU (0.79"). The orbit was solved for the first time in \cite{DeRosa12} and results in a dynamical mass $M_{\mathrm{dyn}} = 4.6 \pm 0.6 M_{\odot}$ using the distance in Table \ref{table:targets}. The primary is a known spectroscopic binary of metallic-line (Am) stars with 0.27 AU separation \citep{Gutmann65}. We obtained interferometric observations of both the primary and secondary. The primary was resolved into the known spectroscopic binary, with masses $1.8 M_{\odot}$ and $1.6 M_{\odot}$ (A8+F0). Meanwhile, the secondary turned out to also be a very close (projected separation 0.06 AU) binary, with masses $0.84 M_{\odot}$ and $0.73 M_{\odot}$ (K1+K4). 

The MSC \citep{Tokovinin18} reports a $0.88 M_{\odot}$ CPM companion at 776" (30900 AU), while \cite{Shaya11} reports a very wide $0.80 M_{\odot}$ CPM companion at 18227" (728308 AU), the latter included as a companion in \cite{DeRosa14}. Both of these companions lie about an order of magnitude above the mutual Keplerian velocity line, and correcting for the proper motion of HIP 80628 by the velocity of its centroid we find that neither of those are true companions. 

\subsection{HIP 80975 ($\omega$ Oph)}

\cite{McDonald12} found $T_{\mathrm{eff}}=7643 \text{ K}$ and $L=34.96 L_{\odot}$ ($R=3.4 R_{\odot}$) for this A star. It is a chemically peculiar (Ap) star with projected rotational velocity $51 \text{ km}\text{ s}^{-1}$ \citep{Abt95}. \cite{DeRosa14} reported a mass $M=2.44 M_{\odot}$ and age $t=450 \text{ Myrs}$. HIP 80975 is located between the upper and lower branches in Fig. \ref{fig:Hip-Gaia} and has no known companion. 

\begin{figure*}
 \includegraphics[width=2.2\columnwidth]{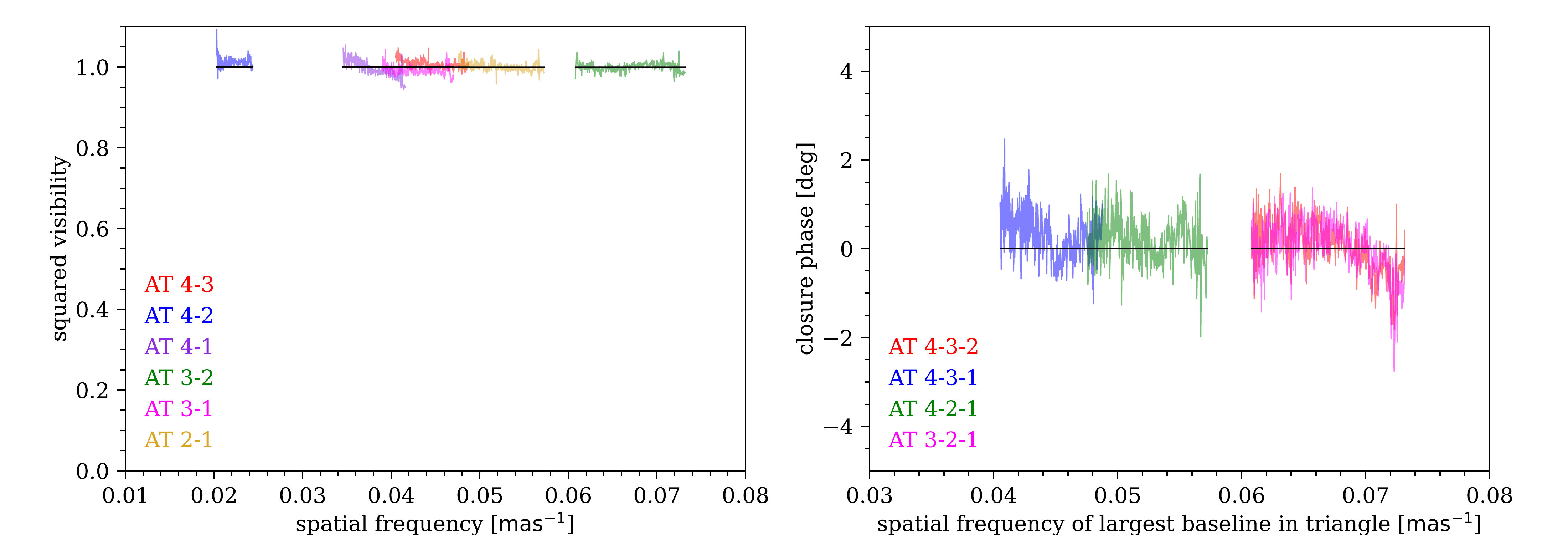}
 \caption{Inteferometric data (colored) for HIP 80975. The black lines show the expected signatures for a single star model.}
 \label{fig:model_fit_80975}
\end{figure*}

\begin{figure}
 \includegraphics[width=\columnwidth]{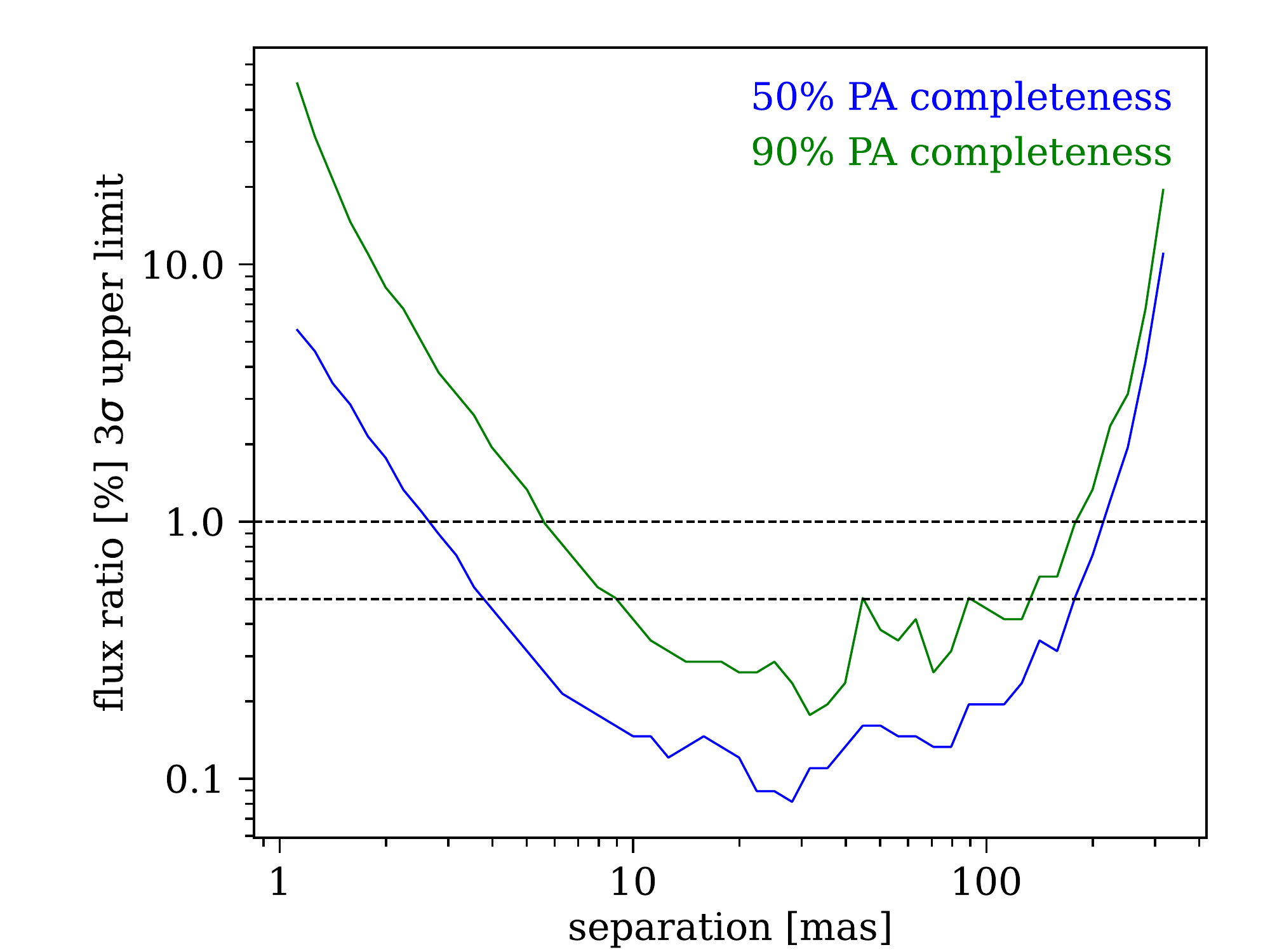}
 \caption{Flux ratio upper limit (3$\sigma$) for a companion in HIP 80975 based on the squared visibilities and closure phases.}
 \label{fig:flux_limit_80975}
\end{figure}

The interferometric data, shown in Fig. \ref{fig:model_fit_80975}, shows no evidence for any companion. Fig. \ref{fig:flux_limit_80975} shows the K band flux ratio upper limits for a companion based on both the squared visibilities and closure phases. Based on the 90\% completeness curve, we can exclude a companion with flux ratio above 0.5\% for projected separations $10 \text{ mas} \leftrightarrow 0.51 \text{ AU} \lesssim \rho \lesssim 130 \text{ mas} \leftrightarrow 6.7 \text{ AU}$ and above 1\% for $\rho \gtrsim 5 \text{ mas} \leftrightarrow 0.26 \text{ AU}$ and $\rho \lesssim 180 \text{ mas} \leftrightarrow 9.2 \text{ AU}$. These flux ratio limits correspond to absolute magnitudes $M_{K} \gtrsim 6.34$ and $M_{K} \gtrsim 5.59$, which translate to masses $M_2 \lesssim 0.38 M_{\odot}$ and $M_2 \lesssim 0.48 M_{\odot}$ according to Eqn. \ref{eqn:Kband_to_mass}. Alternatively, the proper motion changes could be due to a white dwarf companion.  Further constraints on the nature of the companion could be achieved with the upcoming \textit{Gaia} DR3 release based on the acceleration of the A star. 

We note that HIP 80975 was detected in pointed observations with the PSPCs aboard ROSAT \citep{ROSAT00} with a positional error of 2.5". The count rate of $0.055 \pm 0.005 \text{ photons} \text{ s}^{-1}$ within $0.1-2.4$ keV was converted to an X-ray luminosity $L_X = 1.1 \times 10^{29} \text{ erg} \text{ s}^{-1}$ in the catalogue of X-ray emitting A stars of \cite{Schroder07}. Such emission could be explained either by a low mass M dwarf companion that has escaped detection \citep{Magaudda20} or by a relatively hot white dwarf companion.  

\subsection{HIP 87813 / HJ 2814}

We have presented a very detailed analysis of this system in \cite{Waisberg22}. In summary, HIP 87813 is a hierarchical triple consisting of a close (0.16 AU) $2.2 M_{\odot} + 0.85 M_{\odot}$ (A0V+K1V) binary with a $0.74 M_{\odot}$ (K4V) companion on a 24 AU (around 60 yrs) orbit. Furthermore, the triple has a further $1.1 M_{\odot} + 0.9 M_{\odot}$ (G0V+K0V) binary (1.9 AU) companion at a large projected separation of 1600 AU \citep{Tokovinin19} which was missed by \cite{DeRosa14}. The entire system (HJ 2814) is therefore a quintuple. We note that the AO companion at 1.9" discovered by \cite{DeRosa14} is an unrelated background source (Table \ref{table:bogus_companions}). 

\subsection{HIP 95077} 

\cite{McDonald12} found $T_{\mathrm{eff}}=7187 \text{ K}$ and $L=13.50 L_{\odot}$ ($R=2.4 R_{\odot}$) for this A star. \cite{DeRosa14} reported a mass $M=1.87 M_{\odot}$ and age $t=890 \text{ Myrs}$. It has a projected rotational velocity $v \sin i = 78 \text{ km}\text{ s}^{-1}$ \citep{Abt95}. 

\begin{figure}
 \includegraphics[width=\columnwidth]{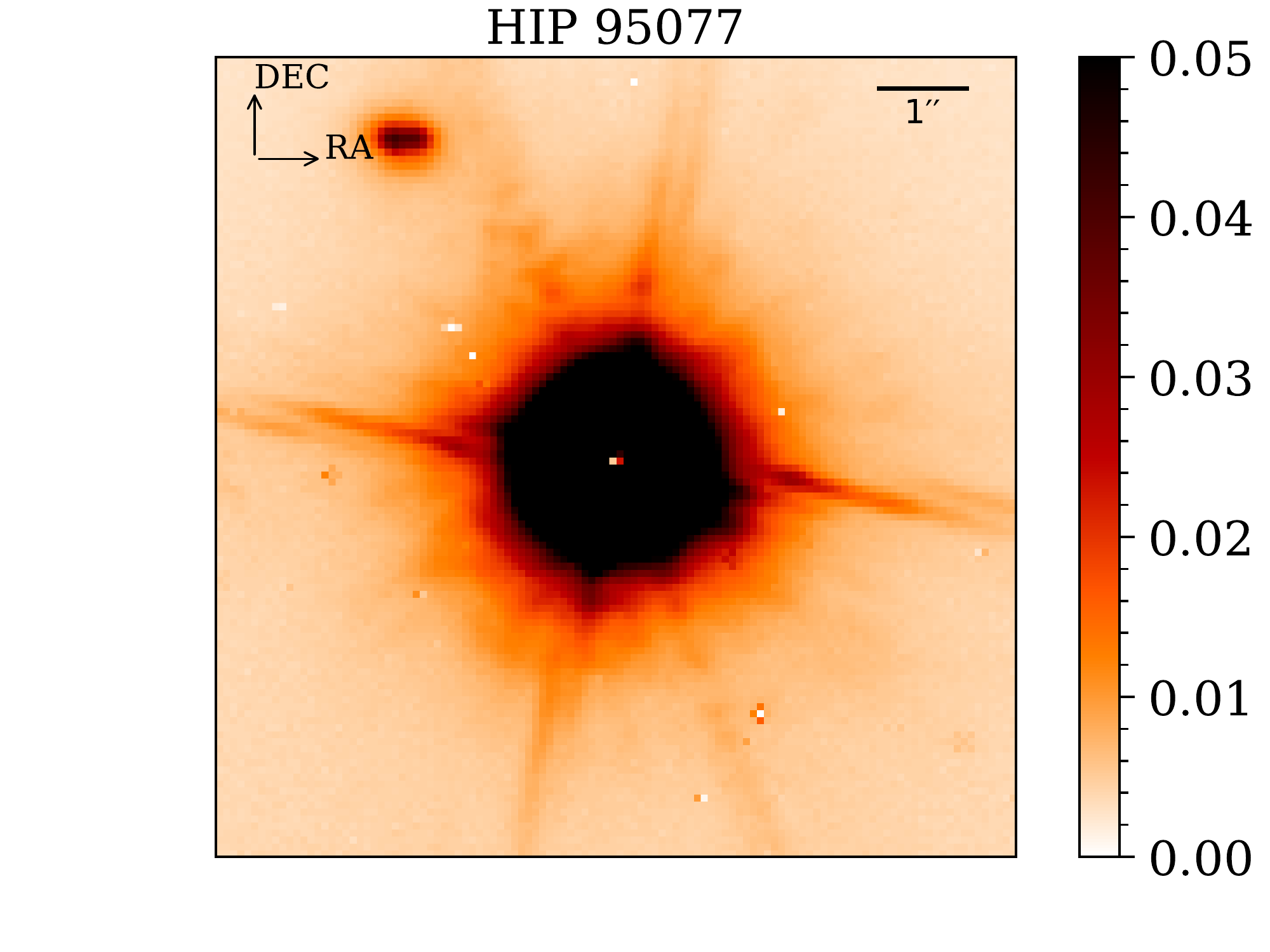}
 \caption{GRAVITY acquisition camera image of HIP 95077 showing the resolved and previously known AO binary companion.}
 \label{fig:acqcam_95077}
\end{figure}

\cite{DeRosa14} also found a new faint AO binary companion to HIP 95077 at projected separations $4.67"$ and $4.74"$ and contrasts $\Delta K=7.07$ and $6.98$, corresponding to masses of $0.15$ and $0.16 M_{\odot}$ if physically associated with the A star. We also see this faint binary companion in the GRAVITY acquisition camera image (Fig. \ref{fig:acqcam_95077}), with the two faint components barely resolved. The position of their centroid relative to the A star is $\Delta \alpha* = -2629 \pm 93 \text{ mas}$ and $\Delta \delta = 3917 \pm 70 \text{ mas}$. This is consistent with the position of their centroid in the 27-06-2008 observation reported in \cite{DeRosa14} ($\Delta \alpha* = -2751 \text{ mas}$ and $\Delta \delta = 3812 \text{ mas}$). Given the high proper motion of the A star (e.g. it moves by $\Delta \delta \approx 400 \text{ mas}$ in 13 years), we conclude that the faint AO companions are not background sources and are almost certainly physically associated to the A star. 

\begin{figure*}
 \includegraphics[width=2.2\columnwidth]{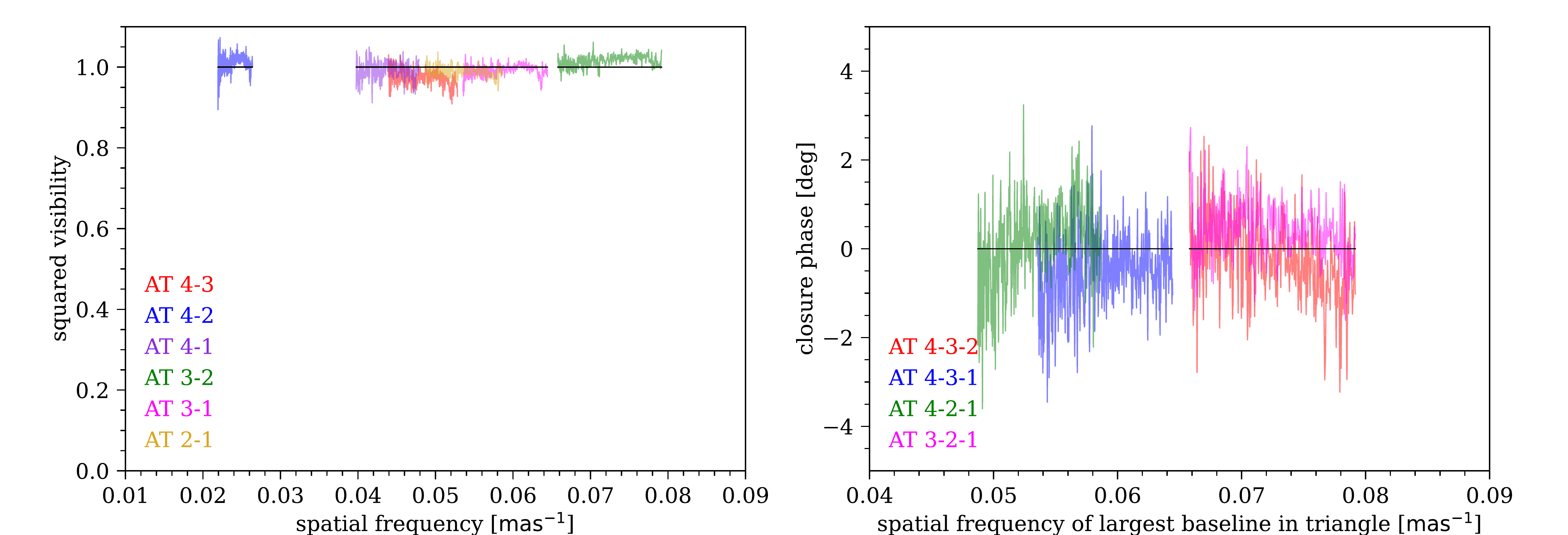}
 \caption{Inteferometric data (colored) for HIP 95077. The black lines show the expected signatures for a single star model.}
 \label{fig:model_fit_95077}
\end{figure*}

\begin{figure}
 \includegraphics[width=\columnwidth]{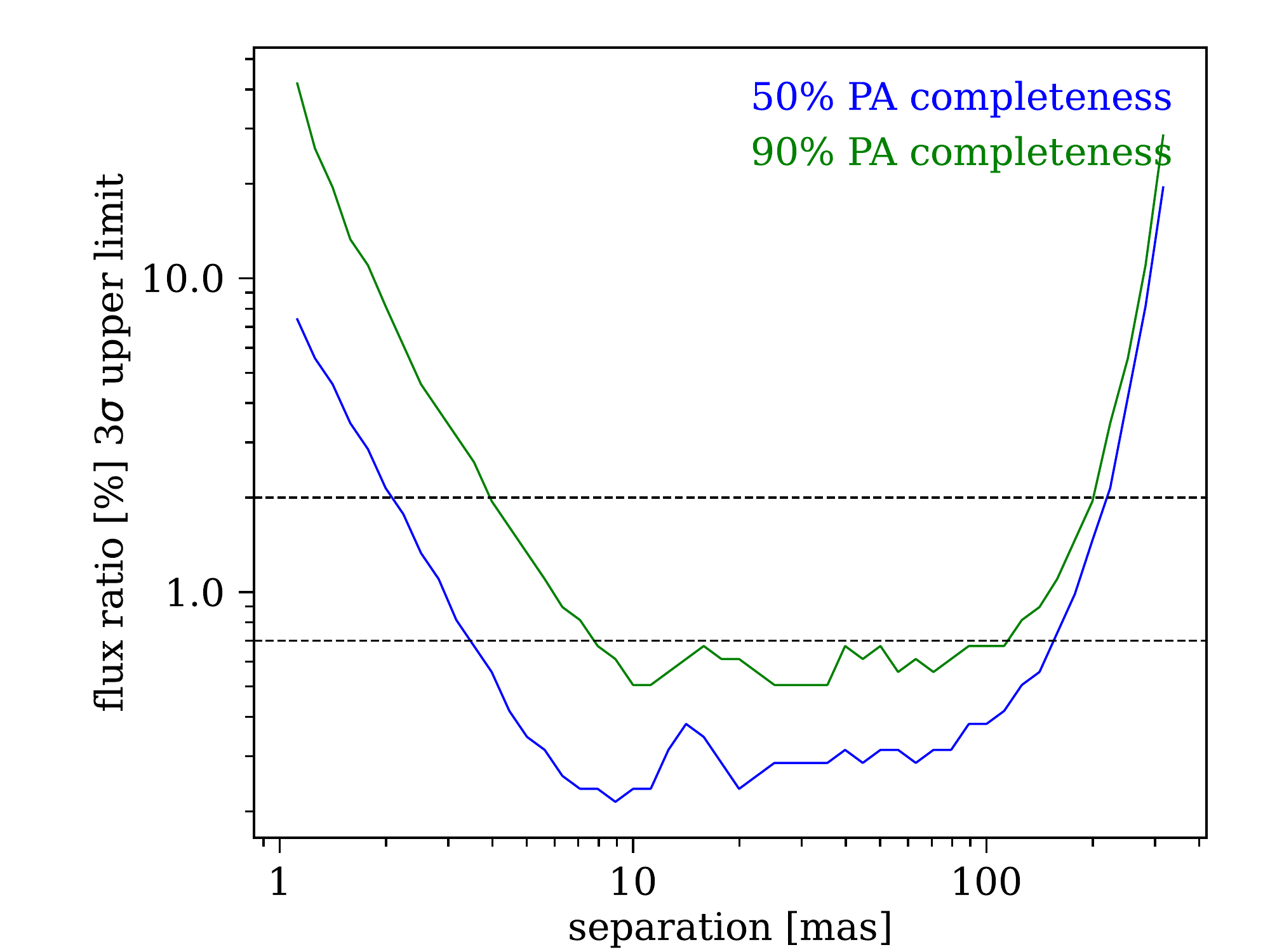}
 \caption{Flux ratio upper limit (3$\sigma$) for a companion in HIP 95077 based on the squared visibilities and closure phases.}
 \label{fig:flux_limit_95077}
\end{figure}

HIP 95077 is located in the upper branch in Fig. \ref{fig:Hip-Gaia} and we therefore expect a companion with a period of a few decades (semi-major axis on the order of 20 AU). We note that the very faint AO binary companions are too far away to explain the proper motion change. However, the interferometric data, shown in Fig. \ref{fig:model_fit_95077}, shows no evidence for any companion. Fig. \ref{fig:flux_limit_95077} shows the K band flux ratio upper limits for a companion based on both the squared visibilities and closure phases. Based on the 90\% completeness curve, we can exclude a companion with flux ratio above 0.7\% for projected separations $8 \text{ mas} \leftrightarrow 0.45 \text{ AU} \lesssim \rho \lesssim 120 \text{ mas} \leftrightarrow 6.8 \text{ AU}$ and above 2\% for $\rho \gtrsim 4 \text{ mas} \leftrightarrow 0.23 \text{ AU}$ and $\rho \lesssim 200 \text{ mas} \leftrightarrow 11.3 \text{ AU}$. These flux ratio limits correspond to absolute magnitudes $M_{K} \gtrsim 6.55$ and $M_{K} \gtrsim 5.41$, which translate to masses $M_2 \lesssim 0.35 M_{\odot}$ and $M_2 \lesssim 0.51 M_{\odot}$ according to Eqn. \ref{eqn:Kband_to_mass}. Alternatively, the proper motion changes could be due to a white dwarf companion.  Further constraints on the nature of the companion could be achieved with the upcoming \textit{Gaia} DR3 release based on the acceleration of the A star. 

\subsection{HIP 104139 ($\theta$ Cap)} 

\cite{Zorec12} found $T_{\mathrm{eff}}=9484\pm153 \text{ K}$, $L=60.9 \pm 2.3 L_{\odot}$ ($R=2.9 R_{\odot}$), $M=2.55\pm0.02 M_{\odot}$ and fractional age on the MS $t_{\mathrm{MS}}=0.715\pm0.034$. It has spectral type A1V and projected rotational velocity $v \sin i = 91 \text{ km}\text{ s}^{-1}$ \citep{Royer07}. \cite{DeRosa14} reported $M=2.68 M_{\odot}$ and age $t=320 \text{ Myrs}$. A companion has been long suspected due to RV variations $\sim 5 \text{ km}\text{ s}^{-1}$ \citep{Lagrange09} but no companion was found either through imaging \citep{Ehrenreich10} or spectroscopy \citep{Gullikson16}. 

\begin{figure*}
 \includegraphics[width=2.2\columnwidth]{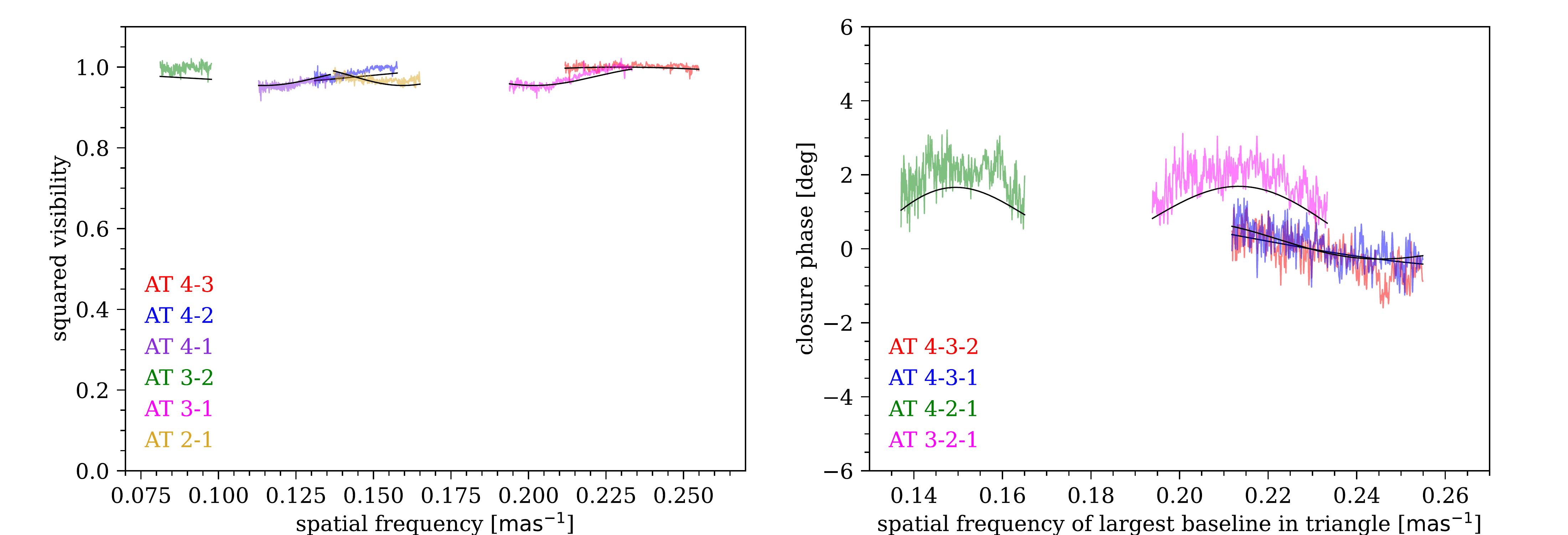}
 \caption{Inteferometric data (colored) and best-fit binary model (black) for HIP 104139.}
 \label{fig:model_fit_104139}
\end{figure*}

HIP 104139 is located at the lower branch in Fig. \ref{fig:Hip-Gaia}, which is yet another evidence for a companion on an orbit of order one year. Indeed, the interferometric data revealed a very faint companion with K band flux ratio 0.012 and projected separation $\rho = 17.0 \text{ mas} \leftrightarrow 0.78 \text{ AU}$. Fig. \ref{fig:model_fit_104139} shows the data and best-fit binary model in black, whose parameters are reported in Table \ref{table:fit_results}. 

From the flux ratio we calculate an absolute K band magnitude of 5.61 for the companion, which corresponds to a mass $M_2=0.48 M_{\odot}$ according to Eqn. \ref{eqn:Kband_to_mass}. In this case, the companion is too faint to affect the properties inferred for the A star. Therefore, this system is an B9.5V+M1.5V binary. The projected separation and the total mass of $3.2 M_{\odot}$ translate to a period estimate of 141 days from Kepler's Third Law. 
\subsection{HIP 116758 ($\omega^1$ Aqr)}

\cite{Zorec12} found $T_{\mathrm{eff}}=7516\pm52 \text{ K}$, $L=17.5 \pm 1.8 L_{\odot}$ ($R=2.5 R_{\odot}$), $M=1.88\pm0.04 M_{\odot}$ and fractional age on the MS $t_{\mathrm{MS}}=0.736\pm0.037$. It has spectral type A7IV and projected rotational velocity $v \sin i = 105 \text{ km}\text{ s}^{-1}$ \citep{Royer07}. \cite{DeRosa14} reported similar values of $M=1.95 M_{\odot}$ and age $t=790 \text{ Myrs}$. This star had no known companion prior to our observations. 

\begin{figure*}
 \includegraphics[width=2.2\columnwidth]{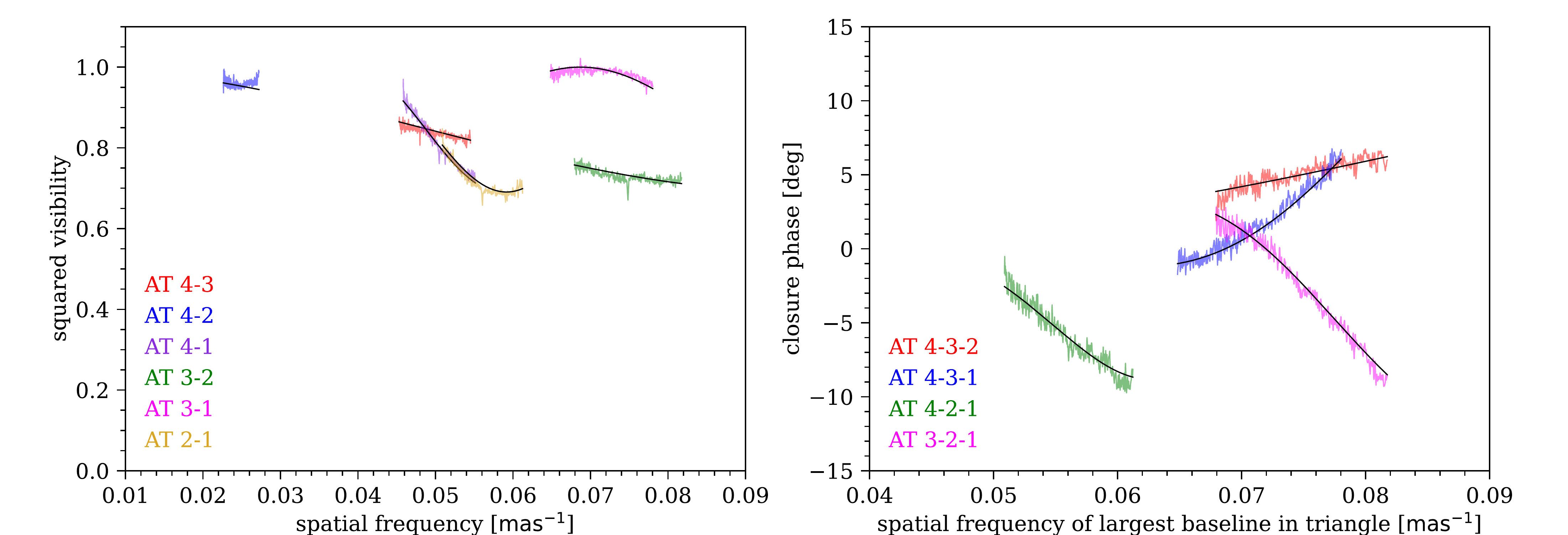}
 \caption{Inteferometric data (colored) and best-fit binary model (black) for HIP 116758.}
 \label{fig:model_fit_116758}
\end{figure*}

HIP 116758 is located at the lower branch in Fig. \ref{fig:Hip-Gaia}; therefore, we expected a companion with an orbital period on the order of 1 year. Indeed, the interferometric data can be perfectly described by a binary model with a K band flux ratio of 0.095 and projected separation $\rho = 26.2 \text{ mas} \leftrightarrow 1.05 \text{ AU}$ (Figure \ref{fig:model_fit_116758}). The best fit results and uncertanties are reported in Table \ref{table:fit_results}. 

From the flux ratio we calculate absolute K band magnitudes of 1.44 and 4.00 for the two components. Using Eqn. \ref{eqn:Kband_to_mass}, we find $M_2 = 0.82 M_{\odot}$. In this case, the companion is massive enough to very slightly bias the properties of the A star inferred as though it were single. The absolute visual magnitude $M_{V_T} = 2.00$ is negligibly affected but $\Delta M_K \approx 0.1$. Using Figs. A.3 and A.5 in \cite{DeRosa14}, we find $M_1 = 1.9 M_{\odot}$ and an age $t = 630 \text{ Myrs}$. This is therefore an A3V+K2V binary. The projected separation and the total mass of $2.7 M_{\odot}$ translate to a period estimate of 239 days from Kepler's Third Law. 

\subsection{HIP 118092} 

\cite{McDonald12} found $T_{\mathrm{eff}}=8598 \text{ K}$ and $L=14.73 L_{\odot}$ ($R=1.7 R_{\odot}$) for this A star. \cite{DeRosa14} reported the discovery of a new AO companion with contrast $\Delta K=3.05$ ($M_2 = 0.71 M_{\odot}$) at a projected separation of 0.35". For the A star, they report a mass $M_1=2.04 M_{\odot}$ and age $t=140 \text{ Myrs}$. We note that the GRAVITY acquisition camera image for this target was saturated because the attenuation filter was not used due to its apparent K band magnitude being above 5; therefore, the image PSF has a FWHM of about 1" and no companion can be seen in the image. 

\begin{figure*}
 \includegraphics[width=2.2\columnwidth]{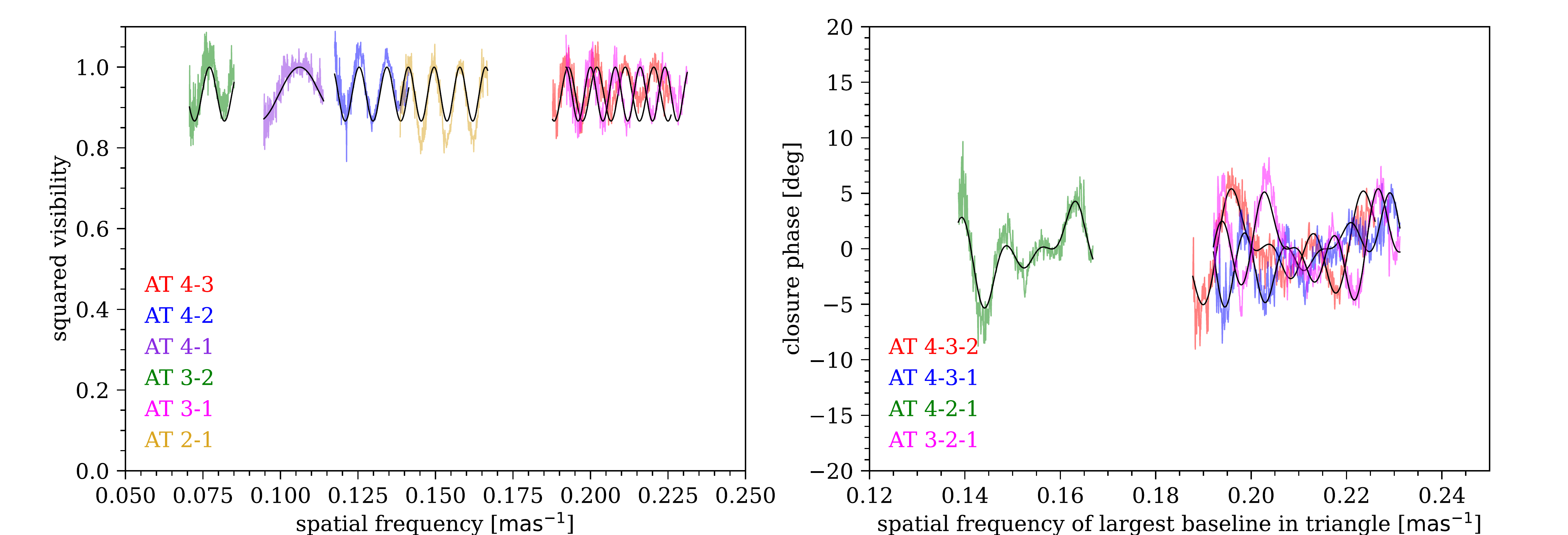}
 \caption{Inteferometric data (colored) and best-fit binary model (black) for HIP 118092.}
 \label{fig:model_fit_118092}
\end{figure*}

HIP 118092 is located at the upper branch in Fig. \ref{fig:Hip-Gaia}, which can be explained by the known AO companion since its projected separation translates to an orbit with a period of decades. The interferometric data reveals the presence of a companion with a K band flux ratio of 0.066 at a projected separation $\rho = 125.0 \text{ mas} \leftrightarrow 7.98 \text{ AU}$. Fig. \ref{fig:model_fit_118092} shows the data and best-fit binary model in black, whose parameters are reported in Table \ref{table:fit_results}. 

The projected separation is quite different from the one reported in \cite{DeRosa14} from a 2006 observation, but the consistent flux ratios confirms that they are the same companion. The change in projected separation by a factor of three points to either a high inclination, high eccentricity or both; further observations would be needed to constrain the orbit. From the flux ratio we calculate an absolute K band magnitude of 4.70 for the companion, which translates into a mass $M_2=0.65 M_{\odot}$. This is therefore a very young A1V+K7V binary system. 

Finally, we note that this star was detected as an X-ray source in the ROSAT all-sky survey \citep{Voges99} with a positional error of 9". The count rate of $0.077 \pm 0.019 \text{ photons} \text{ s}^{-1}$ within $0.1-2.4$ keV was converted to an X-ray luminosity $L_X = 3.1 \times 10^{29} \text{ erg} \text{ s}^{-1}$ in the catalogue of X-ray emitting A stars of \cite{Schroder07}. The emission can be fully explained by the lower mass companion given the youth of the system \citep{Garces11}. 

\section{Results of CPM search} 

Our search resulted in 31 CPM companions within $10^5 \text{ AU}$ to 29 of the 108 A stars in our parent sample (2 of the stars have two companions). 21 of these companions had been previously reported and we list their properties in Table \ref{table:CPM_known}. Several of them were first identified in the AO observations presented in \cite{DeRosa14}, and we therefore confirm them as physical companions. On the other hand, we could exclude the faint AO companion to HIP 109667 and the wide CPM companions to HIP 85922 and HIP 106786 discovered in \cite{DeRosa14} as physical companions based on their very inconsistent parallaxes with the A star (details in Table \ref{table:bogus_companions}). 

\begin{table*}
\centering
\caption{\label{table:CPM_known} Identified CPM companions that were already known among the 108 parent sample of A stars.}
\begin{tabular}{cccccc}
\hline \hline
HIP & \textit{Gaia} eDR3 ID of companion & \shortstack{$G$\\$B_p - R_p$} & \shortstack{Spectral type\\Mass ($M_{\odot}$)} & projected separation (AU) & notes  \\[0.3cm]

3277 & 4908316289753600896 & \shortstack{6.00\\1.09} & \shortstack{K2V\\0.80} & 887 & reported in \cite{DeRosa14} \\[0.3cm]

15353 & 4671503818962008064 & \shortstack{8.81\\1.23} & \shortstack{M1V\\0.50} & 301 & reported in \cite{DeRosa14} \\[0.3cm]

23296 & 3225408155967161216 & \shortstack{13.74\\2.46} & \shortstack{M6V\\0.11} & 452 & reported in \cite{DeRosa14} \\[0.3cm]

26309 & 2907308172059100544 & \shortstack{3.32\\0.55} & \shortstack{F5V\\1.31} & 70885 & reported in \cite{DeRosa14} \\[0.3cm]

29711 & 3020549795181489920 & \shortstack{5.92\\0.90} & \shortstack{K2V\\0.81} & 313 & reported in \cite{DeRosa14} \\[0.3cm]

31167 & 3103325046315850624 & \shortstack{8.73\\1.63} & \shortstack{M1V\\0.51} & 202 & reported in \cite{DeRosa14} \\[0.3cm]

41081 & 5320716335102626048 & \shortstack{2.31\\0.18} & \shortstack{A7V\\1.79} & 80791 & reported in \cite{DeRosa14} \\[0.3cm]

41375 & 3066498145585607424 & \shortstack{11.83\\2.56} & \shortstack{M4.5V\\0.19} & 496 & discovered in \cite{DeRosa14} \\[0.3cm]

48763 & 5658833409227824896 & \shortstack{6.01\\0.76} & \shortstack{K2V\\0.80} & 242 & reported in \cite{DeRosa14} \\[0.3cm]

57013 & 5379417091947347840 & \shortstack{9.18\\2.14} & \shortstack{M2V\\0.45} & 535 & discovered in \cite{DeRosa14} \\[0.3cm]

61498 & 6147117722735170176 & \shortstack{7.71\\2.29} & \shortstack{K8V\\0.62} & 535 & reported in \cite{DeRosa14} \\[0.3cm]

61498 & 6147119548096085376 & \shortstack{8.68\\3.05} & \shortstack{K8V\\0.51} & 12370 & reported in MSC \citep{Tokovinin18} \\[0.3cm]

69995 & 6117782859883553792 & \shortstack{8.57\\0.71} & \shortstack{M0.5V\\0.53} & 263 & discovered in \cite{DeRosa14} \\[0.3cm]

76996 & 5768115519181467264 & \shortstack{6.46\\1.18} & \shortstack{K4V\\0.74} & 4342 & reported in \cite{DeRosa14} \\[0.3cm]

87813 & 4145362250760008832 & \shortstack{4.66\\0.86} & \shortstack{G2V\\1.00} & 1616 & \shortstack{reported in MSC \citep{Tokovinin18}\\ known 1.9 AU spectroscopic binary  \citep{Tokovinin19}} \\[0.3cm]

88726 & 6724105660828668032 & \shortstack{2.46\\0.27} & \shortstack{F0V\\1.65} & 78 & reported in \cite{DeRosa14} \\[0.3cm]

97421 & 6448575090626093440 & \shortstack{5.42\\1.13} & \shortstack{K0V\\0.89} & 5281 & \shortstack{discovered in \cite{DeRosa14}\\very large astrometric noise (probably a binary)} \\[0.3cm]

97423 & 4182644246127569664 & \shortstack{10.03\\0.93} & \shortstack{M3V\\0.37} & 291 & discovered in \cite{DeRosa14} \\[0.3cm]

106654 & 6814117867399713792 & \shortstack{10.73\\-0.29} & \shortstack{WD\\?} & 994 & reported in MSC \citep{Tokovinin18} \\[0.3cm]

107302 & 6838704699743649024 & \shortstack{8.87\\1.37} & \shortstack{M1V\\0.49} & 230 & discovered in \cite{DeRosa14} \\[0.3cm]

117452 & 2328250334633289856 & \shortstack{5.61\\0.98} & \shortstack{K1V\\0.87} & 3287 & reported in MSC \citep{Tokovinin18} \\[0.3cm]

\hline
\end{tabular}
\end{table*}

\begin{figure}
 \includegraphics[width=\columnwidth]{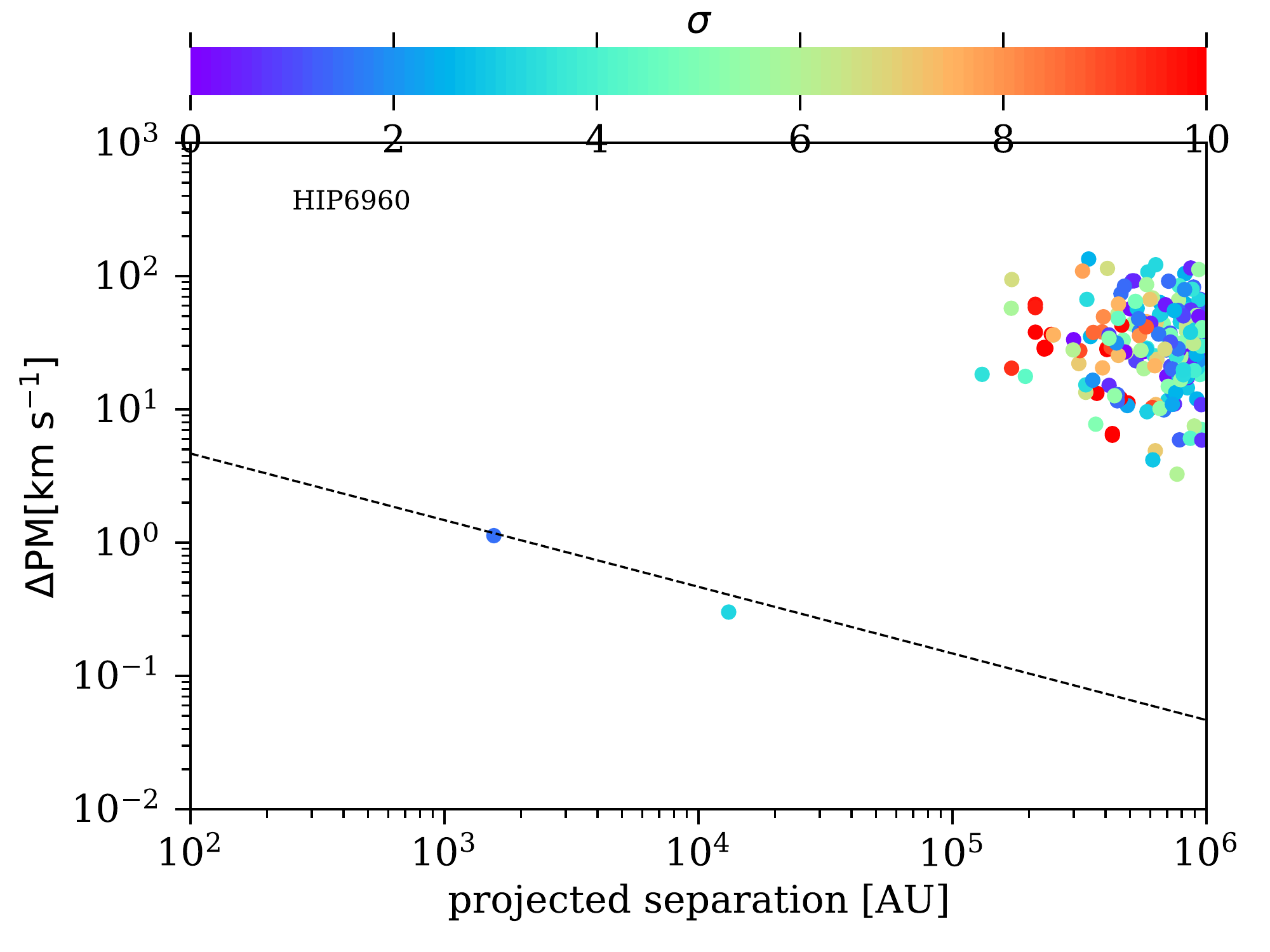}
 \includegraphics[width=\columnwidth]{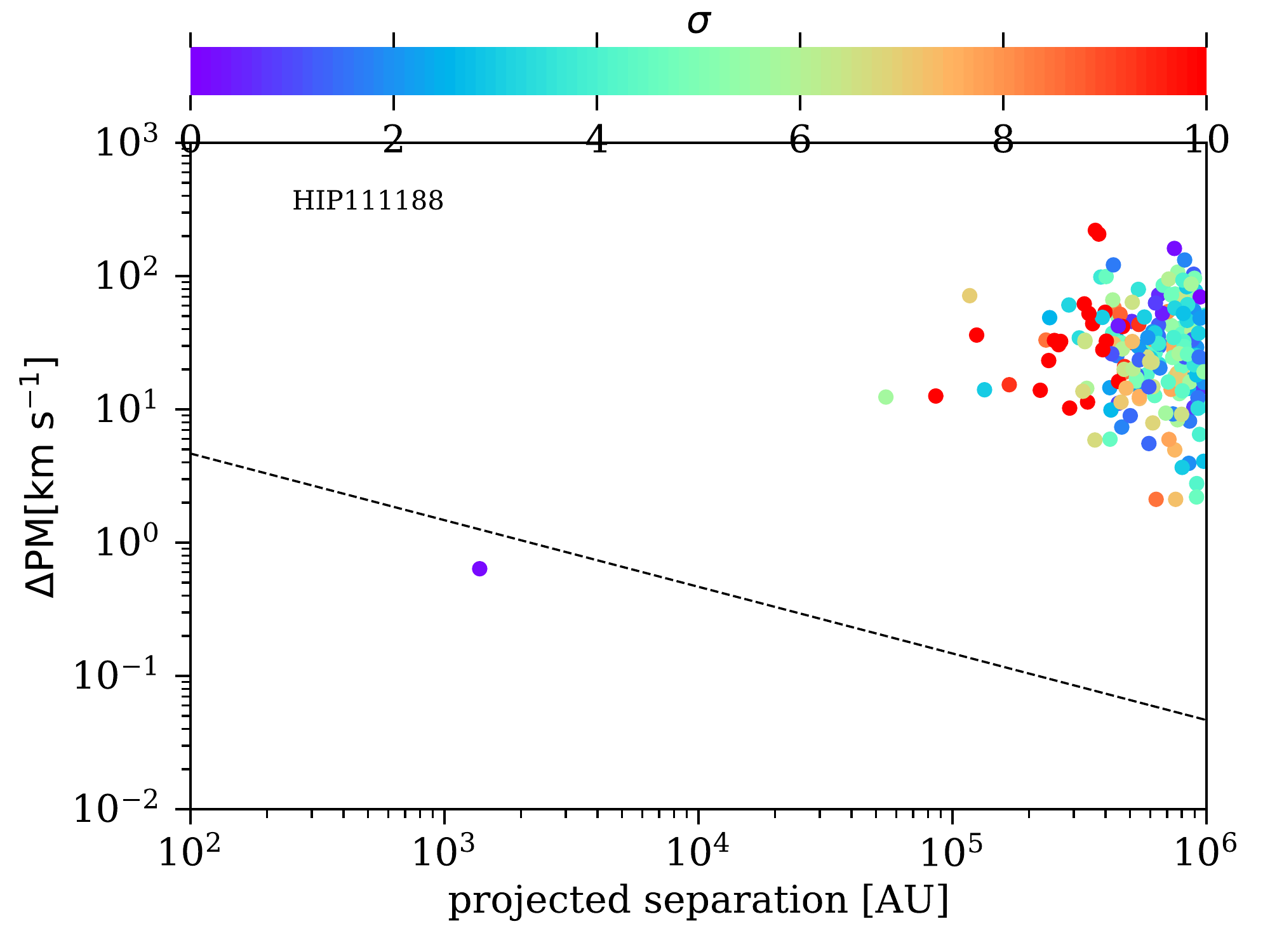}
 \caption{Common proper motion search plots using \textit{Gaia} eDR3 showing two newly found companions for HIP 6960 and one for HIP 111188. $\sigma$ refers to the parallax difference significance. The dashed line shows the expected mutual Keplerian velocity for a total mass of $3 M_{\odot}$.}
 \label{fig:CPM}
\end{figure}

Furthermore, we found 10 CPM companions for which we found no previous reference (either a WDS number or reported in \cite{DeRosa14} or in the MSC \citep{Tokovinin18}), and we list their properties in Table \ref{table:CPM_new}. As is the case for the previously known companions, most of them are low mass M stars, but three of them are relatively massive ($0.66 M_{\odot}$, $0.85 M_{\odot}$ and $1.04 M_{\odot}$). Figure \ref{fig:CPM} shows two examples of our CPM search plots highlighting the two companions ($0.66 M_{\odot}$ and $0.11 M_{\odot}$) to HIP 6960 and our most massive new CPM companion ($1.04 M_{\odot}$) to HIP 111188. In these plots, $\sigma$ refers to the significance in the parallax difference relative to the A star and the dashed line shows the expected mutual Keplerian velocity for a total mass of $3 M_{\odot}$.

\begin{table*}
\centering
\caption{\label{table:CPM_new} Newly identified CPM companions among the 108 parent sample of A stars.}
\begin{tabular}{cccccc}
\hline \hline
HIP & \textit{Gaia} eDR3 ID of companion & \shortstack{$G$\\$B_p - R_p$} & \shortstack{Spectral type\\Mass ($M_{\odot}$)} & projected separation (AU) & notes  \\[0.3cm]

6960 & 5043493513448247040 & \shortstack{7.32\\1.62} & \shortstack{K7V\\0.66} & 1565 & - \\[0.3cm]

6960 & 5043490244977252224 & \shortstack{14.04\\4.11} & \shortstack{M6V\\0.11} & 13146 & - \\[0.3cm]

10069 & 5121585502177248896 & \shortstack{10.08\\2.52} & \shortstack{M3V\\0.37} & 3517 & - \\[0.3cm]

14551 & 5071668498910478336 & \shortstack{10.24\\3.01} & \shortstack{M3V\\0.34} & 3146 & parallax formally inconsistent but high RUWE \\[0.3cm]

23554 & 2960561231044409856 & \shortstack{11.55\\3.04} & \shortstack{M4V\\0.21} & 2808 & - \\[0.3cm]

32938 & 5578901662668399872 & \shortstack{12.48\\3.18} & \shortstack{M5V\\0.16} & 1082 & - \\[0.3cm]

62788 & 3495818747166703232 & \shortstack{5.70\\1.07} & \shortstack{K1V\\0.85} & 11661 & companion is a 0.6" binary (WDS 12520-2648) \\[0.3cm]

76106 & 6254553138488495232 & \shortstack{8.89\\2.16} & \shortstack{M1V\\0.49} & 19667 & - \\[0.3cm]

77464 & 4402445970665447680 & \shortstack{11.86\\3.07} & \shortstack{M4.5V\\0.19} & 4221 & - \\[0.3cm]

111188 & 6601750151432831104 & \shortstack{4.43\\0.73} & \shortstack{G1V\\1.04} & 1376 & - \\[0.3cm]

\hline
\end{tabular}
\end{table*}

\section{Discussion}

The main objective of this paper is observational, namely to introduce our survey strategy and report on the detection of new close companions through NIR interferometry (and wide companions through a common proper motion search with \textit{Gaia} eDR3). Although our interferometric survey is not yet complete, it is already clear that the $\textit{Gaia}$-$\textit{Hipparcos}$ proper motion change is extremely efficient and reliable for identifying multiple systems with periods $P \gtrsim 1 \text{ yr}$, and that when combined with a very preliminary acceleration indicator from \textit{Gaia} (namely, the proper motion difference between DR2 and eDR3) we can efficiently separate between systems with periods up to a few years from periods of decades (lower and upper branches in Figure \ref{fig:Hip-Gaia}). In the following, we briefly discuss some implications that can already be inferred from our results, focusing in particular on the connection between A stars and WDs which is our main motivation for the observations. 

\subsection{Updated multiplicity of A stars}
\label{section:multiplicity}

The multiplicity statistics of A stars as a function of companion separation and mass ratio provides crucial observational constraints for star formation models. At this point, however, we refrain from a detailed statistical analysis of the two dimensional (separation and mass ratio) multiplicity function of A stars for two reasons, namely (i) our interferometric survey is not yet complete (ii) the upcoming \textit{Gaia} DR3 release should provide a very complete sample of short-period binaries as well as accelerations in the very near future. Although we did end up discovering some short period ($\sim 0.1 \text{ AU}$) binaries in high multiplicity systems (HIP 5300, HIP 47479, HIP 80628, HIP 87813), the bulk of such binaries are not targeted by our survey. The large astrometric noise of several targets below the dashed line in Figure \ref{fig:Hip-Gaia} suggests at least some short-period binaries are still to be discovered; in particular, only one (HIP 80628) of the seven Am stars in our sample is a known spectroscopic binary, whereas spectroscopic surveys suggest that all \citep{Abt65} or at least most \citep{Carquillat07} Am stars have a very close companion. 

In any case, we can provide lower limits to the overall multiplicity of A stars by compiling all the currently known and confirmed companions to our parent sample of A stars based on the previous AO observations in \cite{DeRosa14}, our newly discovered interferometric companions, our CPM search with \textit{Gaia} eDR3, companions listed in the MSC \citep{Tokovinin18} and in the literature \citep[Table 9 in][]{DeRosa14}\footnote{We note that there are two stars with hints of a companion in the literature, namely HIP 2578 \citep{vandenBos27} and HIP 62983 \citep{Africano75}, for which we found no companion in VLT/NACO images downloaded and reduced from the ESO archive (following the same steps as in \cite{Waisberg22}). In both cases, the suggested separation is 0.1" and the mass ratio around unity, so that the companion should be easily detectable if it existed. Therefore, we do not consider these companions to be real, and they are therefore included in Table \ref{table:bogus_companions}.}. We currently count 74 companions to the 108 A stars in our parent sample, compared to 50 companions reported in \cite{DeRosa14} for these same A stars. We have therefore increased the number of companions by 50\% in our sample, even after discarding 8 of the companions reported in \cite{DeRosa14} (Table \ref{table:bogus_companions}). Out of the 108 A stars, we currently have that 56 (52\%) are single, 37 (34\%) are binaries, 9 (8\%) are triples, 5 (5\%) are quadruples and 1 (1\%) is a quintuple, corresponding to an overall multiplicity fraction of 48\% and a companion fraction of 0.7. They are similar to the numbers reported by \cite{DeRosa14} (except for a higher percentage of quadruples) based on their most complete subset of 156 A stars with both AO observations and CPM search (which are mostly northern targets for which the completeness for spectroscopic binaries is higher). However, we expect a significant increase in our numbers due to unknown spectroscopic binaries, many of which might be discovered in the upcoming \textit{Gaia} DR3 release. 

The separation distribution fitted by \cite{DeRosa14} has a log-normal dependency with a peak at around 400 AU, with a reduction by about a factor of three at 30 AU (the lower limit they adopt for their bins). This is rather surprising for two reasons. Firstly, this peak is significantly farther and sharper than the one found for solar mass stars \citep[which have a broad peak at 10-100 AU][]{Raghavan10}. Secondly, dedicated spectroscopic surveys show that the separation distribution of A stars has a similar value to the peak in the separation distribution found by \cite{DeRosa14} for periods of around 1000 days \citep{Abt65}, corresponding to a separation of around 3 AU for a total mass of $3 M_{\odot}$. 

For our sample of 108 A stars, there were originally seven stars with a companion with a projected separation between 5 and 30 AU known in \cite{DeRosa14}, corresponding to a value of 0.08 for a bin within those separation limits in Figure 9 of \cite{DeRosa14}. We have found two new such companions (HIP 54477 and HIP 87813), and we note that another very low mass companion has been discovered in HIP 23554 within this separation range \citep{DeRosa19}. This gives ten stars, or 0.12 for the value in Figure 9 in \cite{DeRosa14}. Furthermore, if we consider HIP 95077, a star in the upper branch of Figure \ref{fig:Hip-Gaia} for which we found no companion with $M \gtrsim 0.5 M_{\odot}$, as well as the two stars also in the upper branch for which we still do not have interferometric observations but in which we also expect a companion in this separation range (HIP 30666 and HIP 48763; we note that HIP 29852 already has such a known companion), the number of stars increases to 13 and the corresponding value to 0.15. This is already a factor of three higher than what would have been predicted based on the separation distribution of \cite{DeRosa14}, and strongly suggests that the apparent suppression of companions at a few tens of AU is not real but rather a result of observational bias related to the difficulty of detecting such systems. We have shown that high precision \textit{Hipparcos}-\textit{Gaia} astrometry is an excellent way to uncover these systems. 

\subsection{A stars and white dwarfs} 

\begin{figure*}
 \includegraphics[width=\columnwidth,angle =90]{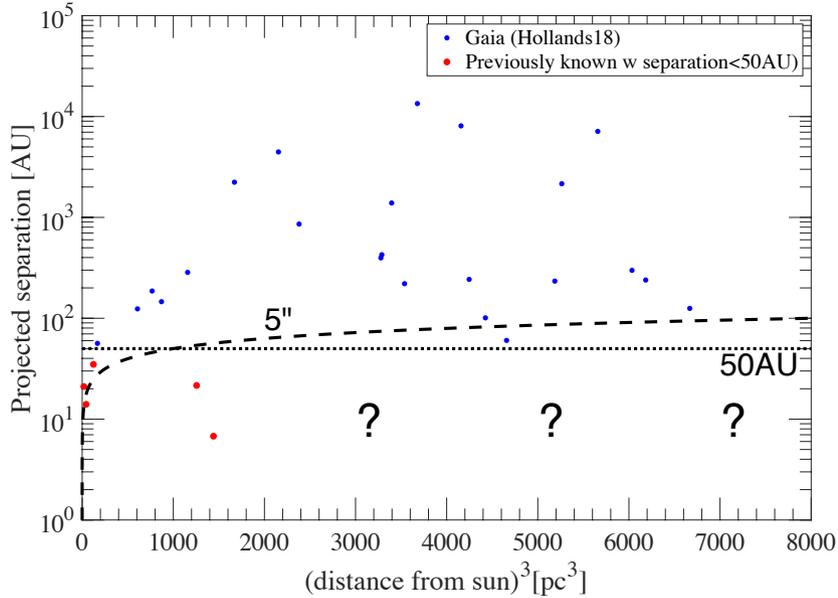}
 \caption{The sample of known white dwarfs with a known companion within 20 pc of the Sun. The dashed line marks the 5" limit below which \textit{Gaia} is not complete, while the dotted horizontal line marks the 50 AU separation below which there is a clear lack of systems beyond about 10 pc. For the red points only, the separation refers to the physical semi-major axis of the orbit.}
 \label{fig:WD_20pc}
\end{figure*}

Fig. \ref{fig:WD_20pc} plots the sample of known WDs within 20 pc of the Sun and which have a known
companion, as a function of their cubed distance and the binary projected separation. The points in blue are wide binaries resolved by \textit{Gaia} \citep{Hollands18}, and are near or above the limit $\rho > 5"$ for which \textit{Gaia} cannot detect a faint WD amidst the glow of a much brighter MS companion. On the other hand, we plot in red the five WDs with known companions with semi-major axis $< 50 \text{ AU}$ (whose properties we summarize in Table \ref{table:nearby_WDs}). Three of them are the very nearby WDs Sirius B, Procyon B and 40 Eridani B, and the two farther points are Gliese 86B and G107-70. From Fig. \ref{fig:WD_20pc}, it is clear that a significant population of WDs within 20 pc with MS companions $\lesssim 50 \text{ AU}$ has escaped detection. In particular, there are four WDs with MS companions with (projected) separation $\lesssim 60 \text{ AU}$  (the three nearby systems mentioned above plus Stein 2051 B) within 5.5 pc; if that is not a statistical fluke, we would expect around 200 similar systems within 20 pc but only one additional such system is known (Gliese 86B; we note that the companion to G107-70 is another WD). 

\begin{table*}
\centering
\caption{\label{table:nearby_WDs} Properties of known WDs within 20 pc with companions at $\lesssim 50$ AU}
\begin{tabular}{cccccc}
\hline \hline
WD & WD mass ($M_{\odot}$) & Companion mass ($M_{\odot}$) & distance (pc) & semi-major axis (AU) & References \\[0.3cm]
Sirius B & $1.018\pm0.011$ & $2.063\pm0.023$ & 2.6 & 20 & [1] \\[0.3cm]
Procyon B & $0.602\pm0.015$ & $1.499\pm0.031$ & 3.5 & 15 & [2,3,4] \\[0.3cm]
40 Eridani B & $0.573\pm0.018$ & $0.84$ & 5.0 & 35 & [5,6] \\[0.3cm]
Stein 2051 B & $0.675\pm0.051$ & $0.252\pm0.013$ & 5.5 & 55 & [7,8,9] \\[0.3cm]
Gliese 86B & $0.595\pm0.010$ & $0.83 \pm 0.05$ & 10.8 & 22 & [10] \\[0.3cm]
G107-70 & $\sim 0.5$ & $\sim 0.5$ (WD) & 11.3 & 7 & [11] \\[0.3cm]
\hline
\multicolumn{6}{l}{[1] \cite{Bond17}; [2] \cite{Provencal02}; [3] \cite{Liebert13}; [4] \cite{Bond15}; [5] \cite{Nordstrom04}} \\
\multicolumn{6}{l}{[6] \cite{Mason17}; [7] \cite{Strand89}; [8] \cite{Sahu17} ; [9] \cite{Khata20}; [10] \cite{Brandt19}} \\
\multicolumn{6}{l}{[11] \cite{Harrington81}} \\
\end{tabular}
\end{table*}

\subsubsection{Where are the Procyon progenitors?}
\label{section:Procyon} 

Another remarkable fact is that these same four closest WDs in Table \ref{table:nearby_WDs} are also among the six closest overall WDs to the Sun. Again, assuming this is not a statistical fluke, this would imply that quite a large fraction of WD progenitors should have a companion with separation $1 \text{ AU} < a < 30 \text {AU}$ (since the mass loss during the formation of the WD widens the orbit somewhat). Three of these four WDs (except Sirius B) have masses consistent with having been born as A stars. One of the two isolated WDs within the six nearest WDs (van Maanen's Star) also has a mass consistent with an A star progenitor, while GJ 440 likely descended from a B star. Therefore, out of the four nearest WDs that could have had an A star progenitor, three of them have a MS companion with a separation of a few tens of AU and in two of them the companion has a mass higher than $0.8 M_{\odot}$ (Procyon B and 40 Eridani B). Taken at face value, this would suggest that a very large fraction of A stars (50\%) should have such a massive companion at these separations. 

Our selection based on \textit{Gaia}-\textit{Hipparcos} proper motion changes is nearly fully complete for Procyon and 40 Eridani progenitors within our parent sample of 108 A stars. There are currently only eight A stars with a $M > 0.8 M_{\odot}$ companion at a separation (projected or semi-major axis, if the orbit is known) $1 \text{ AU} \lesssim a_{\mathrm{proj}} \lesssim 30 \text{ AU}$, namely HIP 5300, HIP 29852, HIP 47479, HIP 69974, HIP 77660, HIP 80628, HIP 106786 and HIP 116758 (for $M > 0.7 M_{\odot}$, there are only two additional stars). There are a further five stars that we still have not observed with intetferometry which could have such a companion. Therefore, we can constrain the fraction of Procyon progenitors among A stars to between 7\% and 12\%. This is much lower than what would have been inferred based on the very nearby (< 6pc) WDs, and we therefore conclude the latter are simply statistical anomalies. 

Another interesting fact is that out of the eight A stars listed above, four are now known to have a high (3+) multiplicity structure (HIP 5300, HIP 47479, HIP 77660 and HIP 80628), and this number could increase with future interferometric observations for the remaining targets. This implies that in a large fraction of the Procyon progenitors, there will be nontrivial evolution once the more massive star evolves off the main sequence due to the presence of a very close companion.

Finally, we note that we are almost surely still missing several WDs with MS companions closer than 50 AU within 20 pc \citep{Katz14}. In our parent sample of 108 A stars, there are currently 14 stars with confirmed companions (any mass) with projected separations $1 \text{ AU} \lesssim a_{\mathrm{proj}} \lesssim 30 \text{ AU}$, which give a lower limit of 13\% for their fraction. Given that there are 140 WDs within 20 pc, we would then expect about 18 WDs hidden within the glow of a MS companion still to be discovered within this distance. Such WDs are particularly challenging to detect, as there are even single WDs as close as 13 pc that have only been detected recently \citep{Hollands18}. 

\subsubsection{Current A star - WD  binaries}

There are 636 photometrically identified A stars within 75 pc \citep{DeRosa14}; therefore, one would expect to find the first A star at a distance of about 9 pc. The fact that Sirius A is located at 2.6 pc is therefore already a statistical anomaly, and we can show that it is even more so given its WD companion Sirius B. For this we use our sample to put an upper limit on the relative number of A stars that currently have a WD companion within our detectability limits.

On the one hand, out of the 108 A stars in our parent sample, there is only one with a confirmed WD companion, namely HIP 106654 with a CPM WD companion at a projected separation of 994 AU. For our target distances of 40-75 pc, our CPM search with \textit{Gaia} should be complete to WDs with absolute \textit{Gaia} g magnitudes above 13.0-11.6 (assuming a maximum apparent magnitude of 21) and for separations greater than 200-375 AU (for a minimum angular separation of 5"). Therefore, we estimate that a fraction on the order of $1\%$ of A stars have such a WD companion. 

On the other hand, our selection based on a \textit{Gaia}-\textit{Hipparcos} proper motion change greater than $0.5 \text{ km}\text{ s}^{-1}$ is 90\% complete for $M \gtrsim 1 M_{\odot}$ with projected separation $a_{\mathrm{proj}} < 20 \text{ AU}$ and for periods larger than a couple decades. The completeness is not a strong function of the mass, however, and for $M \gtrsim 0.5 M_{\odot}$ our completeness falls to about 80\%. Furthermore, we are also sensitive to high and low mass companions on shorter period orbits with projected separation around 1 AU, as the lower branch in Figure \ref{fig:Hip-Gaia} shows. 

There are some stars with proper motion changes for which we could not detect the culprit companion in our interferometric observations, namely HIP 70931, HIP 77660, HIP 80975 and HIP 95077 down to a limiting mass of about $0.5 M_{\odot}$ for a MS star. Furthermore, there are six stars within our sample for which interferometric observations have still not been obtained. By assuming that these are all WDs, we can put a very generous upper limit of about $10\%$ to the fraction of A stars with a WD companion within a separation $1 \text{ AU} \lesssim a \lesssim 20 \text{ AU}$. We expect this limit to be significantly more constraining with the upcoming \textit{Gaia} DR3 release (which could provide orbits or accelerations for these systems, which will constrain the mass of the companion) and the completion of our interferometric survey. In any case, there is no doubt that Sirius is a very extreme anomaly among A stars. 
Together with the anomalous high density of nearby WDs with (massive) MS companions discussed in Section \ref{section:Procyon}, this suggests an unusually high density of intermediate mass stars in the relatively recent past in the very close vicinity to the Sun. 

\section*{Acknowledgements}

This work has made use of data from the European Space Agency (ESA) mission Gaia (https://www.cosmos.esa.int/gaia), processed by the Gaia Data Processing and Analysis Consortium (DPAC, https://www.cosmos.esa.int/web/gaia/dpac/consortium). Funding for the DPAC has been provided by national institutions, in particular the institutions participating in the Gaia Multilateral Agreement. This publication makes use of data products from the Two Micron All Sky Survey, which is a joint project of the University of Massachusetts and the Infrared Processing and Analysis Center/California Institute of Technology, funded by the National Aeronautics and Space Administration and the National Science Foundation. This research has made use of the Jean-Marie Mariotti Center \texttt{SearchCal} service \footnote{Available at http://www.jmmc.fr/searchcal} co-developped by LAGRANGE and IPAG. This reaseach has made use of the CDS Astronomical Databases SIMBAD and VIZIER \footnote{Available at http://cdsweb.u-strasbg.fr/},  NASA's Astrophysics Data System Bibliographic Services, NumPy \citep{van2011numpy} and matplotlib, a Python library for publication quality graphics \citep{Hunter2007}.

\section*{Data Availability}

All the data used in this paper is (or will soon become) publicly available from the respective archives.

\bibliographystyle{mnras}
\bibliography{main} 


\appendix

\onecolumn

\section{Discarded companions} 

\begin{table}
\centering
\caption{\label{table:bogus_companions} Companions in \citep{DeRosa14} identified as not physical among our parent sample of 108 A stars.}
\begin{tabular}{cccc}
\hline \hline
HIP & companion type & separation & reason \\[0.3cm]

2578 & visual/AO companion & 0.1" & no evidence in NACO image from 2004-11-20 \\[0.3cm]

54746 & visual companion & 3.80" & apparent mistake in Table 9 \\[0.3cm]

62983 & visual/AO companion & 0.1" & no evidence in NACO image from 2005-02-06 \\[0.3cm]

80628 & CPM companion & 18227" & not a companion based on \textit{Gaia} \\[0.3cm]

85922 & CPM companion & 135.6" & inconsistent parallaxes (21.4 vs 7.5 mas) \\[0.3cm]

87813 & AO companion & 1.88" & background source \citep{Waisberg22} \\[0.3cm]

106786 & CPM companion & 375.6" & inconsistent parallaxes (18.2 vs 7.0 mas) \\[0.3cm]

109667 & AO companion & 5.14" & inconsistent parallaxes (15.7 vs 1.8 mas) \\[0.3cm]

\hline
\end{tabular}
\end{table}


\bsp	
\label{lastpage}
\end{document}